\documentclass[journal]{IEEEtran}

\usepackage{graphicx}
\usepackage{epstopdf}
\usepackage{color}
\usepackage{amsfonts}
\usepackage{amssymb}
\usepackage{amsmath} 
\usepackage{cite}
\usepackage{epsfig}
\usepackage[tight,footnotesize]{subfigure}

\newtheorem{thm}{\hspace{-0.5cm} \bf{Theorem}}[section]
\newtheorem{lem}[thm]{\hspace{-0.5cm} \bf{Lemma}}
\newtheorem{defn}[thm]{\hspace{-0.5cm} \bf{Definition}}
\newtheorem{assumption}[thm]{\hspace{-0.5cm} \bf{Assumption}}

\begin{document}

\title{Performance Analysis of Joint Time Delay and Doppler-Stretch Estimation with Random Stepped-Frequency Signals}

\author{Tong~Zhao, Zheng~Nan, and Tianyao~Huang
\thanks{Tong~Zhao, Zheng~Nan, and Tianyao~Huang are with the RF Tech. R\&D Lab., AVIC Beijing Keveen Aviation Instrument Co., Ltd..}
}

\maketitle

\begin{abstract}

This paper investigates the performance of joint time delay and Doppler-stretch estimation with the random stepped-frequency (RSF) signal. Applying the ambiguity function (AF) to implement the estimation, we derive the compact expressions of the theoretical mean square errors (MSEs) under high signal-to-noise ratios (SNRs). The obtained MSEs are shown consistent with the corresponding Cramer-Rao lower bounds (CRLBs), implying that the AF-based estimation is approximately efficient. 
Waveform parameters including higher carrier frequencies, wider bandwidth covered by the carrier frequencies, and frequency shifting codewords with larger variance are expected for a better estimation performance.
As a synthetic wideband signal, the RSF signal achieves the same estimation performance as the OFDM signal within an identical bandwidth. Due to its instantaneous narrowband character, requirement for the bandwidth of the receiver is much reduced.

\end{abstract}


\IEEEpeerreviewmaketitle

\section{Introduction}

The stepped-frequency (SF) signal has been widely adopted in modern wideband radar and sonar systems. Compared to the conventional narrowband signal, the SF signal achieves higher range resolution, and the multiple scatterers of the target can be thoroughly distinguished. Since its energy is dispersed to the whole bandwidth covered by the carrier frequencies, the SF signal attains a lower probability of interception \cite{Axelsson07}. Meanwhile, classified as one of the synthetic wideband signals, the SF signal only takes up a narrow bandwidth at any time instant, while the whole carrier frequency bandwidth can be occupied if an instantaneous wideband signal (e.g. the OFDM signal) is employed. From this scope, the SF signal could largely reduce the requirements for the bandwidth of the receiver \cite{Skolnik}.

The linear stepped-frequency (LSF) signal uses a fixed frequency shifting step, which introduces a ``ridge'' in its ambiguity function (AF). This causes a coupling problem between the range and Doppler dimensions \cite{Huang14}. For the random stepped-frequency (RSF) signal, however, with the carrier frequencies of the pulses randomly distributed over a given bandwidth, its AF appears in a thumbtack alike shape, where the range and Doppler dimensions are completely decoupled \cite{Huang12}.
The resolutions in both dimensions thus meet further improvements, and the range ambiguity is efficiently suppressed.
Besides, as the frequency shifting codeword of the RSF signal is usually highly self-correlated and hard to track, the interference between adjacent radars can be largely reduced, whereas the electronic counter-countermeasures (ECCM) capabilities can be also acquired \cite{Huang14}. As a result, parameter estimation with the RSF signal is of great practical significance and is gaining increasing research interests.


The joint estimation of time delay and Doppler-stretch is a fundamental problem that facilitates target tracking and localization in radar and sonar systems, upon which the location and velocity information of the target is managed to be attained \cite{Niu99}. One of the standard methods for joint delay-Doppler estimation is to adopt the AF \cite{Auslander84, Lush91}. By locating the peak of the AF, the joint estimation was primarily implemented \cite{Luo95}.
So as to evaluate the performance of the estimation, the Cramer-Rao lower bound (CRLB) is commonly employed, for the reason that it is regarded as a theoretical lower bound for the variance of any unbiased estimation and is usually easy to calculate. Based on a wideband signal model, the CRLBs of time delay and Doppler-stretch were derived, under the assumption that the scattering coefficient of the target was known a priori \cite{Luo95}. Derivation for a more realistic case was performed by \cite{Wei07}, where the scattering coefficient was supposed to be unavailable at the receiver. The estimation problem of an extended target with multiple scatterers was even considered in \cite{Zhao15}.
Nevertheless, the CRLB is only reliable for accurately presenting the estimation performance when it is approached by the corresponding mean square error (MSE), i.e. when the estimation is (asymptotically) \emph{efficient} \cite{Kay}. By directly calculating the theoretical MSE, the estimation performance can be straightforwardly revealed. However, in most cases it is difficult to evaluate the MSE in a theoretical manner. Among the limited number of related works, the MSEs of the AF-based joint estimation were calculated in \cite{Luo95}, whereas a correction followed in \cite{Yin98}. However, the calculating results are not accurate since too many approximations were made in both works.

All the approaches \cite{Luo95, Yin98, Wei07, Zhao15} were built on a general wideband signal model, which failed to reveal the waveform parameters that influence the estimation performance with any specific signal. Based on the LSF signal model, the CRLBs of time delay and Doppler-stretch were derived in \cite{Li11}, whereas those of the high resolution range (HRR) profiles of an extended target were provided by \cite{Liu14}. Similar range and Doppler estimation problems with the OFDM signal were also considered in \cite{Sen14, Gu15, Turlapaty14}.
However, performance analysis on the parameter estimation with the RSF signal is quite limited in previous works.

In this paper, we investigate the performance of the AF-based joint delay-Dppler estimation with the RSF signal. Under high signal-to-noise ratio (SNR) assumption, compact expressions of the MSEs are obtained through a novel and strict derivation. The MSEs are shown consistent with their corresponding CRLBs, revealing that the AF-based estimation is approximately efficient. As illustrated by the derivations and simulations, three waveform parameters of the RSF signal, namely, the central carrier frequency, the bandwidth covered by the carrier frequencies, and the variance of the frequency shifting codeword mainly influence the estimation performance. In specific, by increasing either the bandwidth for the carrier frequencies or the variance of the frequency shifting codeword the performance of delay estimation can be improved, while better performance of Doppler-stretch estimation calls for higher central carrier frequency.
As one of the synthetic wideband signals, the RSF signal only takes up a narrow bandwidth at any time instant, whereas it achieves the similar estimation performance as the OFDM signal does, contributing to a much reduced requirement for the bandwidth of the receiver.

The rest of the paper is organized as follows. Section~II describes the model of the RSF signal and makes necessary preliminaries. Section~III gives the main results of this paper, while Section~IV provides the derivations for the main results. Section~V verifies the main results by numerical examples, followed by a conclusion in Section~VI.

\section{Modeling and Preliminaries}

Consider an RSF signal with $K$ pulses. Let $\beta(t)$ denote the envelope of each pulse. The transmitted signal is then modeled as
\begin{equation}\label{equ:16}
s(t) = \sum_{k = 0} ^{K - 1} \beta(t - k T_r) e^{j 2 \pi f_k (t - k T_r)},
\end{equation}
where $f_k$ denotes the carrier frequency of the $k$-th pulse, and $T_r$ is the pulse repetition interval (PRI). Assume that the carrier frequency remains constant within each pulse and shifts randomly over the pulses within a given bandwidth. Then the carrier frequencies of the pulses can be represented as
\begin{equation}
f_k = f_0 + d_k \delta_f, \quad k = 0, 1, \ldots, K - 1,
\end{equation}
where $f_0$ denotes the central carrier frequency, $\delta_f$ is the minimum frequency shifting step, $[ d_0, d_1, \ldots, d_{K - 1} ]^T$ is the frequency shifting codeword, in which $d_k$ for each pulse is randomly selected from $[-M, M]$, $M \in \mathbb{N^+}$. Suppose that a target is moving along the line of sight (LOS) with a constant radial velocity $v$ relative to the sensor (e.g. a radar or a sonar). The echo reflected from the moving target is then given by \cite{Luo95}
\begin{equation}\label{equ:echo}
s_r(t) = x s(\gamma_0 (t - \tau_0)),
\end{equation}
where $x$ is the scattering coefficient of the target, accounting for the attenuation and reflection, $\tau_0$ and $\gamma_0 = \frac{c - v}{c + v}$ represent the time delay and Doppler-stretch, respectively, with the wave propagation velocity denoted as $c$.
The signal received by the sensor is contaminated by a white Gaussian noise (WGN) $w(t)$ with power spectral density $N_0$. The received signal is then expressed as
\begin{equation}\label{equ:19}
\begin{aligned}
y(t) 
& = x s(\gamma_0 (t - \tau_0)) + w(t).
\end{aligned}
\end{equation}
Sampled at the rate of $1 / \Delta$, the received signal \eqref{equ:19} turns into
\begin{equation}
y(n \Delta) = x s(\gamma_0 (n \Delta - \tau_0)) + w(n \Delta),
\end{equation}
$n = 0, 1, \ldots, N-1$, where $N$ is the total number of sampling points, $w(n \Delta)$ is distributed as $\mathbb{C} N (0, \sigma^2)$, and $\sigma^2 \Delta = N_0$ \cite{Kay}.

\subsection{Ambiguity Function}

As mentioned in Section~I, the AF
\begin{equation}\label{equ:21}
A_{y s}(\tau, \gamma) = \sum_{n = 0}^{N - 1} y(n \Delta) s^*(\gamma (n \Delta - \tau))
\end{equation}
is generally applied for implementation of the joint delay-Doppler estimation \cite{Luo95, Kay}. 
Denoted as
\begin{equation}\label{equ:0063}
(\hat{\tau}, \hat{\gamma}) = \arg \max_{ \tau \in [\tau_{\text{min}}, \tau_{\text{max}}], \gamma \in [\gamma_{\text{min}}, \gamma_{\text{max}}] } \left | A_{y s}(\tau, \gamma) \right |,
\end{equation}
the AF-based estimation is implemented by locating the peak of the AF \cite{Luo95, Yin98},
where $\hat{\tau}$ and $\hat{\gamma}$ respectively denote the estimations of time delay and Doppler-stretch, and $[\tau_{\text{min}}, \tau_{\text{max}}] \times [\gamma_{\text{min}}, \gamma_{\text{max}}]$ regulates the searching area for the estimations of the parameters. Note that both $\hat{\tau}$ and $\hat{\gamma}$ are continuous random variables that vary with $\sigma$. In addition, we also assume $(\tau_0, \gamma_0) \in [\tau_{\text{min}}, \tau_{\text{max}}] \times [\gamma_{\text{min}}, \gamma_{\text{max}}]$.

\subsection{Preliminaries}

For the ease of problem statement and derivations in the following sections, we make the necessary assumptions and definitions.
%
%


%
%
\begin{assumption}\label{assumption:04}{\it
The envelope of each pulse $\beta(t)$ is time-limited within $[0, T]$, $T < \frac{T_r}{2}$.
}\end{assumption}

Assumption~\ref{assumption:04} guarantees that each pulse does not overlap with any other ones in time domain.
%
%
\begin{assumption}{\it
$s \in \mathbb{C}^{(2)}(\mathbb{R})$, i.e. $s: \mathbb{R} \rightarrow \mathbb{R}$ has continuous derivatives up to order 2 inclusive.
}\end{assumption}
\begin{assumption}\label{assumption:2.8}
\begin{equation}
\begin{aligned}
& \textstyle \frac{\partial^2}{\partial \tau^2} \left | A_{s_r s} \right |^2 (\tau_0, \gamma_0) \cdot \frac{\partial^2}{\partial \gamma^2} \left | A_{s_r s} \right |^2 (\tau_0, \gamma_0) - \\
& \textstyle \left ( \frac{\partial^2}{\partial \tau \partial \gamma} \left | A_{s_r s} \right |^2 (\tau_0, \gamma_0) \right )^2 \neq 0.
\end{aligned}
\end{equation}
\end{assumption}
\begin{assumption}\label{assumption:01}{\it
$ | A_{s_r s}(\tau, \gamma) | $ has the unique maximizer $(\tau_0, \gamma_0)$, which holds almost surely if the sampling rate $1 / \Delta$ is greater than the bandwidth of $s(t)$.
}\end{assumption}
\begin{defn}\label{defn:2.6}{\it
For any sequences $\{ a_k; k = 0, 1, \ldots, K - 1 \}$ and $\{ b_k; k = 0, 1, \ldots, K - 1 \}$, define}
\begin{align}
\nonumber & \text{Var}\left \{ a_k \right \} := \left ( \text{Std}\left \{ a_k \right \} \right )^2 := \frac{1}{K} \sum_{k = 0}^{K - 1} a_k^2 - \left ( \frac{1}{K} \sum_{k = 0}^{K - 1} a_k \right )^2, \\
\nonumber & \text{Cov}\left \{ a_k, b_k \right \} := \frac{1}{K} \sum_{k = 0}^{K - 1} a_k b_k - \frac{1}{K} \sum_{k = 0}^{K - 1} a_k \frac{1}{K} \sum_{k = 0}^{K - 1} b_k, \\
& \rho(a_k, b_k) := \frac{ \text{Cov}\left \{ a_k, b_k \right \} }{ \text{Std}\left \{ a_k \right \} \text{Std}\left \{ b_k \right \} }.
\end{align}
\end{defn}
\begin{assumption}\label{assumption:2.7}{\it
For RSF signals,}
%
%
\begin{equation}
\text{Cov}\left \{ k^i, d_k^j \right \} = 0, \quad \forall i, j = 1, 2.
\end{equation}
\end{assumption}

\begin{defn}{\it $\forall i = 0, 2$,}
\begin{equation}
\begin{aligned}
& \textstyle S_i^{(0)} := \int_{0}^{T} \left ( t - \frac{T}{2} - \text{Std} \left \{ T_k \right \} \right )^i \beta^{2}(t) dt, \\
& \textstyle S_i^{(1)} := \int_{0}^{T} \left ( t - \frac{T}{2} - \text{Std} \left \{ T_k \right \} \right )^i \dot{\beta}^{2}(t) dt,
\end{aligned}
\end{equation}
{\it where $T_k := k T_r + \frac{T}{2}$, and $\dot{\beta}(t) := \frac{d}{d t} \beta(t)$.}
\end{defn}

\section{Main Results}

In this section, the main results of the paper are provided.

Under high SNRs, i.e. when $\sigma$ is sufficiently small compared to the amplitude of the echo, the unbiasedness of the AF-based joint estimation is shown as:

\begin{thm}\label{lemma:08}{\it
For each $\epsilon > 0$,
$$
P \left \{ \| (\hat{\tau}, \hat{\gamma}) - (\tau_0, \gamma_0) \|_2 > \epsilon \right \} \rightarrow 0 \text{ as } \sigma \rightarrow 0,
$$
indicating that
\begin{equation}
\hat{\tau} \stackrel{P}{\longrightarrow} \tau_0 \text{ and } \hat{\gamma} \stackrel{P}{\longrightarrow} \gamma_0
\end{equation}
as $\sigma \rightarrow 0$, where $\| (x, y) \|_2 = \left ( x^2 + y^2 \right )^{\frac{1}{2}}$, and ``$x \stackrel{P}{\longrightarrow} y$'' denotes that $x$ converges to $y$ in probability.
}\end{thm}
\begin{IEEEproof}
See Appendix~\ref{proof:theorem01}.
\end{IEEEproof}
%
%

The estimations of time delay and Doppler-stretch respectively converge to their true values as the noise level gets weaker, implying that the AF-based joint estimation is asymptotically {\em unbiased} when SNR is sufficiently large \cite{Trees}.

Building on Theorem~\ref{lemma:08}, we then evaluate the MSEs of the AF-based estimation.
Define
\begin{align}
\nonumber & \begin{array}{ll}
\displaystyle B := \int_{-\infty}^{\infty} \left | \dot{s}(t) \right |^2 dt, & \displaystyle E := \int_{-\infty}^{\infty} \left | s(t) \right |^2 dt, \\
\displaystyle C := \int_{-\infty}^{\infty} t \left | \dot{s}(t) \right |^2 dt, & \displaystyle F := \text{Im} \left \{ \int_{-\infty}^{\infty} s(t) \dot{s}^*(t) dt \right \}, \\
\displaystyle D := \int_{-\infty}^{\infty} t^2 \left | \dot{s}(t) \right |^2 dt, & \displaystyle G := \text{Im} \left \{ \int_{-\infty}^{\infty} s(t) t \dot{s}^*(t) dt \right \},
\end{array} \\
\nonumber & \begin{array}{l}
\displaystyle \Pi := \left ( E B - F^2 \right ) \left ( E D - G^2 \right ) - \left ( E C - F G \right )^2,
\end{array} \\
& \begin{array}{l}
\displaystyle \Pi_0 := \Pi - \frac{5}{4} E^2 (E B - F^2).
\end{array}
\end{align}
The theoretical MSEs of time delay and Doppler-stretch are formally given by:

\begin{thm}\label{theorem:02}{\it
In the AF-based estimation, the theoretical MSEs of time delay and Doppler-stretch satisfy
\begin{subequations}\label{equ:0034}
\begin{align}
\nonumber & \left [ \begin{array}{cc}
\displaystyle \lim_{N_0 \rightarrow 0} \textstyle \mathbb{E} \left \{ \frac{\left | \hat{\tau} - \tau_0 \right |^2}{N_0} \right \},
& \displaystyle \lim_{N_0 \rightarrow 0} \textstyle \mathbb{E} \left \{ \frac{\left | \hat{\gamma} - \gamma_0 \right |^2}{N_0} \right \}
\end{array} \right ]^T \\
\label{equ:0034a} = & \frac{E}{2 |x|^2 \Pi_0} \left [ \begin{aligned}
& \frac{1}{\gamma_0} \left (
\begin{array}{l}
\left ( E D - G^2 - \frac{5}{4} E^2 \right )^2 \left ( E B - F^2 \right ) +  \\
\left ( E C - F G \right )^2 \left ( E D - G^2 + \frac{3}{4} E^2 \right ) - \\
2 \left ( E C - F G \right )^2 \left ( E D - G^2 - \frac{5}{4} E^2 \right )
\end{array}
\right ) \\
& \gamma_0^3 \left (
\begin{array}{l}
\left ( E B - F^2 \right )^2 \left ( E D - G^2 + \frac{3}{4} E^2 \right ) + \\
\left ( E C - F G \right )^2 \left ( E B - F^2 \right ) - \\
2 \left ( E B - F^2 \right ) \left ( E C - F G \right )^2
\end{array}
\right )
\end{aligned} \right ] \\
\label{equ:0034b} \approx & \frac{E}{2 |x|^2 \Pi} \left [ \begin{array}{cc}
\frac{1}{\gamma_0} \left (E D - G^2 \right ),
& \gamma_0^3 \left (E B - F^2 \right )
\end{array} \right ]^T,
\end{align}
\end{subequations}
where the approximate in (\ref{equ:0034b}) holds if $\beta(t) = \beta(T - t)$.
}\end{thm}
\begin{IEEEproof}
See Section~\ref{proof:theorem02}.
\end{IEEEproof}


In Theorem~\ref{theorem:02}, \eqref{equ:0034a} precisely describes the MSEs of the AF-based estimation under high SNRs. If the additional condition is involved, we obtain the approximated, but much simplified forms of the MSEs, which are presented by \eqref{equ:0034b}.
The condition $\beta(t) = \beta(T - t)$ suggests that the envelope $\beta(t)$ is a symmetric function with an axis of symmetry $t = T / 2$.
This is yielded by most of the radar signals. In this sense, Theorem~\ref{theorem:02} reliably presents the MSEs of most cases.
Moreover, note that the MSEs of time delay and Doppler-stretch are represented with integrations, corresponding to the case where the sampling rate is sufficiently large, i.e. $\Delta \rightarrow 0$, such that the summations can be replaced by integrations.

The effectiveness of the AF-based estimation is judged by evaluating the gaps between the MSEs and their corresponding CRLBs, since the latters represent the minimum achievable variances of any unbiased estimation.
In \cite{Luo95} and \cite{Wei07}, the CRLBs for joint delay-Doppler estimation with known and unknown scattering coefficient were respectively derived. In the AF-based joint estimation as introduced in Section~II-A, only the magnitude information of the AF is utilized. However, the phase information contained in the scattering coefficient $x$ is ignored. 
Therefore, regarding $x$ as one of the unknown parameters and according to the calculation of CRLBs in \cite{Wei07, Zhao15}, we have

\begin{thm}\label{theorem:crlb}{\it \footnote{The proof of Theorem~\ref{theorem:crlb} is omitted due to limited space.}
The CRLBs of time delay and Doppler-stretch are given by}
\begin{equation}\label{equ:crlb}
\left [ \begin{array}{c}
\text{CRLB}_{\tau} \\
\text{CRLB}_{\gamma}
\end{array} \right ] = \frac{N_0 E}{2 |x|^2 \Pi} \left [ \begin{aligned}
& \frac{1}{\gamma_0} (E D - G^2) \\
& \gamma_0^3 (E B - F^2)
\end{aligned} \right ].
\end{equation}
%
%
%
%
\end{thm}

Clearly, the MSEs in \eqref{equ:0034b} are confirmed to be consistent with the corresponding CRLBs. By Theorems~\ref{lemma:08}--\ref{theorem:crlb}, we describe the AF-based estimation as an approximately {\em efficient} estimation \cite{Trees}. 

We should also notice that Theorem~\ref{theorem:02} and \ref{theorem:crlb} in fact apply to estimations with arbitrary wideband signals modeled by \eqref{equ:echo} (including the RSF signals). However, the specific waveform parameters determining the estimation performance are not able to be revealed by \eqref{equ:0034} or \eqref{equ:crlb}.
So as to explore the relationship between the estimation performance and the waveform parameters of the RSF signal, we substitute the signal model \eqref{equ:16} into \eqref{equ:0034b} and obtain the compact expressions of the MSEs:


%
%
\begin{thm}\label{theorem:03}{\it For an RSF signal, the theoretical MSEs of joint delay-Doppler estimation under high SNRs are specifically expressed as}
\begin{subequations}\label{equ:theorem3}
\begin{align}
\nonumber & \lim_{N_0 \rightarrow 0} \mathbb{E} \textstyle \left \{ \frac{\left | \hat{\tau} - \tau_0 \right |^2}{N_0} \right \} \\
\label{equ:theorem3:1} & \approx \textstyle \frac{1}{2 \gamma_0 |x|^2 K} \left \{ \left [ S_0^{(1)} + 4 \pi^2 \text{Var}\left \{ f_k \right \} S_0^{(0)} \right ]^{-1} + \right . \\
\nonumber & \quad \textstyle \left . \left ( \frac{1}{K} \sum_{k = 0}^{K - 1} T_k \right )^2 \textstyle \left [ S_2^{(1)} + 4 \pi^2 \left ( \frac{1}{K} \sum_{k = 0}^{K - 1} f_k^2 \right ) S_2^{(0)} \right ]^{-1} \right \}, \\
& \begin{aligned}\label{equ:theorem3:2}
& \lim_{N_0 \rightarrow 0} \textstyle \mathbb{E} \left \{ \frac{\left | \hat{\gamma} - \gamma_0 \right |^2}{N_0} \right \} \\
& \approx \textstyle \frac{\gamma_0^3}{2 |x|^2 K} \left [ S_2^{(1)} + 4 \pi^2 \left ( \frac{1}{K} \sum_{k = 0}^{K - 1} f_k^2 \right ) S_2^{(0)} \right ]^{-1}.
\end{aligned}
\end{align}
\end{subequations}
\end{thm}
\begin{IEEEproof}
See Section~\ref{proof:theorem03}.
\end{IEEEproof}

As revealed by \eqref{equ:theorem3}, the estimation performance with an RSF signal is mainly dominated by three factors, namely, the central carrier frequency, the bandwidth covered by the carrier frequencies, and the frequency shifting pattern. Specifically,
i)~In both the expressions of MSEs above, there exists a component $\frac{1}{K} \sum_{k = 0}^{K - 1} f_k^2$. Since $f_0 \gg \delta_f$, $\frac{1}{K} \sum_{k = 0}^{K - 1} f_k^2 \approx f_0^2$. Therefore, both the performances of time delay and Doppler-stretch estimations can be improved if the central carrier frequency $f_0$ increases.
ii)~The component $\text{Var}\left \{ f_k \right \}$ in \eqref{equ:theorem3:1} can be rewritten as $\delta_f^2 \text{Var}\left \{ d_k \right \}$, where $\delta_f$ determines the available bandwidth for the carrier frequencies, while $\text{Var}\left \{ d_k \right \}$ is related to the frequency shifting pattern.

Beyond the revealings by the expressions of the MSEs in Theorem~\ref{theorem:02}, simulation results in Section~V show that the performance of delay estimation is only slightly improved as the central carrier frequency increases. This indicates that the performance of delay estimation is mainly influenced by the covered bandwidth of the carrier frequencies and the variance of the frequency shifting codeword.

With the analyses above, the theoretical MSEs given by Theorem~\ref{theorem:03} could serve as a guidance for waveform design, which aims at properly configuring the waveform parameters and improving the estimation performance. By increasing the central carrier frequency the performance of Doppler-stretch estimation can be significantly improved, while the performance of delay estimation can be improved by increasing the bandwidth covered by carrier frequencies and by adopting the frequency shifting codewords with large $\text{Var}\left \{ d_k \right \}$.

\subsection{Comparison with Other Waveforms}

The estimation performance with the RSF signal is then fairly compared with those with a monotone signal and an OFDM signal, where the three signals are comprised of the same number of pulses with the same amount of energy. The monotone signal fixes its carrier frequency to $f_0$ for all pulses, whereas the OFDM signal simultaneously uses $L$ orthogonal subcarriers within each pulse.

i) The estimation performance with the monotone signal is directly known from \eqref{equ:theorem3}, since the signal can be considered as a special case of RSF signal with $\delta_f = 0$. The MSEs of time delay and Doppler-stretch then readily reduce to
\begin{subequations}\label{equ:monotone}
\begin{align}
& \begin{aligned}\label{equ:monotone:1}
& \lim_{N_0 \rightarrow 0} \textstyle \mathbb{E} \left \{ \frac{\left | \hat{\tau} - \tau_0 \right |^2}{N_0} \right \} 
\approx \textstyle \frac{1}{2 \gamma_0 |x|^2 K} \left \{ \left ( S_0^{(1)} \right )^{-1} + \right . \\
& \quad \textstyle \left . \left ( \frac{1}{K} \sum_{k = 0}^{K - 1} T_k \right )^2 \textstyle \left [ S_2^{(1)} + 4 \pi^2 f_0^2 S_2^{(0)} \right ]^{-1} \right \},
\end{aligned} \\
\label{equ:monotone:2} & \lim_{N_0 \rightarrow 0} \textstyle \mathbb{E} \left \{ \frac{\left | \hat{\gamma} - \gamma_0 \right |^2}{N_0} \right \} \approx \textstyle \frac{\gamma_0^3}{2 |x|^2 K} \left [ S_2^{(1)} + 4 \pi^2 f_0^2 S_2^{(0)} \right ]^{-1},
\end{align}
\end{subequations}
with $\text{Var} \left \{ f_k \right \} = 0$ and $\frac{1}{K} \sum_{k = 0}^{K - 1} f_k^2 = f_0^2$.
Recalling that $\text{Var} \left \{ f_k \right \}$ mainly determines the MSE of time delay, and that $\text{Var} \left \{ f_k \right \} > 0$ always holds in \eqref{equ:theorem3:1}, we discover that the RSF signal leads to a better performance of delay estimation than the monotone signal.
While due to the reason that $\frac{1}{K} \sum_{k = 0}^{K - 1} f_k^2 \approx f_0^2$ holds in \eqref{equ:theorem3:2}, the MSEs of Doppler-stretch with the two signals are approximately the same.

ii) The OFDM signal can be considered as the sum of $L$ ``monotone signals'' regulated by $1 / \sqrt{L}$, which is denoted as
\begin{equation}\label{equ:ofdm}
s(t) = \frac{1}{\sqrt{L}} \sum_{k = 0}^{K - 1} \sum_{l = 0}^{L - 1} \beta(t - k T_r) e^{j 2 \pi f_l (t - k T_r)},
\end{equation}
where $f_l = f_0 + d_l \delta_f$ \cite{Gu15, Turlapaty14}. All the available carrier frequencies within the given bandwidth for the RSF signal are simultaneously used by the OFDM signal as its subcarriers. Substituting \eqref{equ:ofdm} into \eqref{equ:0034b} and following the derivation in Section~\ref{proof:theorem03}, we obtain
\begin{subequations}\label{equ:mse_ofdm}
\begin{align}
\nonumber & \lim_{N_0 \rightarrow 0} \textstyle \mathbb{E} \left \{ \frac{\left | \hat{\tau} - \tau_0 \right |^2}{N_0} \right \} \\
& \approx \textstyle \frac{1}{2 \gamma_0 |x|^2 K} \left \{ \left [ S_0^{(1)} + 4 \pi^2 \text{Var}\left \{ f_l \right \} S_0^{(0)} \right ]^{-1} + \right . \\
\nonumber & \quad \textstyle \left . \left ( \frac{1}{K} \sum_{k = 0}^{K - 1} T_k \right )^2 \left [ S_2^{(1)} + 4 \pi^2 \left ( \frac{1}{L} \sum_{l = 0}^{L - 1} f_l^2 \right ) S_2^{(0)} \right ]^{-1} \right \}, \\
\label{equ:mse_ofdm:2} & \begin{aligned}
& \lim_{N_0 \rightarrow 0} \textstyle \mathbb{E} \left \{ \frac{\left | \hat{\gamma} - \gamma_0 \right |^2}{N_0} \right \} \\
& \approx \textstyle \frac{\gamma_0^3}{2 |x|^2 K} \left [ S_2^{(1)} + 4 \pi^2 \left ( \frac{1}{L} \sum_{l = 0}^{L - 1} f_l^2 \right ) S_2^{(0)} \right ]^{-1}.
\end{aligned}
\end{align}
\end{subequations}
%
%
By comparing \eqref{equ:mse_ofdm} with \eqref{equ:theorem3}, it is obvious that the estimation performances with the OFDM and RSF signals are the same under high SNRs. Nevertheless, as one of the synthetic wideband signals, the RSF signal only takes up a narrow instantaneous bandwidth, while on the contrary, the whole bandwidth is occupied by all the subcarriers when transmitting an OFDM signal. From this perspective, the requirements for the receiver can be much reduced if the RSF signal is employed for parameter estimation.

Summarily, as for the performance of delay estimation, the RSF and OFDM signals have the identical performance under high SNRs. Due to their wideband character, they both outperform the narrowband monotone signal. While for the performance of Doppler-stretch estimation, which is mostly dependent on the central carrier frequency, the three signals have approximately the same performance.

\section{Derivations and Proofs}

In this section, we provide the detailed derivations of the main results. 

Since the AF-based joint estimation can be interpreted as searching for the maximum point of the AF, we start our derivation by focusing on the properties of $\left | A_{y s}(\tau, \gamma) \right |^2$ at its maximum point. Obviously, the partial derivatives of $\left | A_{y s}(\tau, \gamma) \right |^2$ with respect to $\tau$ and $\gamma$ both reach zero at $(\hat{\tau}, \hat{\gamma})$ \cite{Luo95, Yin98}, i.e.,
\begin{equation}\label{equ:013}
\begin{aligned}
& \textstyle \frac{\partial}{\partial \tau} \left | A_{y s} \right |^2 (\hat{\tau}, \hat{\gamma}) = 2 \text{Re} \left \{ A^*_{y s}(\hat{\tau}, \hat{\gamma}) \frac{\partial}{\partial \tau} A_{y s}(\hat{\tau}, \hat{\gamma}) \right \} = 0, \\
& \textstyle \frac{\partial}{\partial \gamma} \left | A_{y s} \right |^2 (\hat{\tau}, \hat{\gamma}) = 2 \text{Re} \left \{ A^*_{y s}(\hat{\tau}, \hat{\gamma}) \frac{\partial}{\partial \gamma} A_{y s}(\hat{\tau}, \hat{\gamma}) \right \} = 0.
\end{aligned}
\end{equation}
Letting
\begin{align}
\nonumber & \textstyle A_{s_r s}(\tau, \gamma) := \sum_{n = 0}^{N - 1} x s(\gamma_0 (n \Delta - \tau_0)) s^*(\gamma (n \Delta - \tau)), \\
& \textstyle A_{n s}(\tau, \gamma) := \sum_{n = 0}^{N - 1} w(n \Delta) s^*(\gamma (n \Delta - \tau)),
\end{align}
we have $A_{y s}(\hat{\tau}, \hat{\gamma}) = A_{s_r s}(\hat{\tau}, \hat{\gamma}) + A_{n s}(\hat{\tau}, \hat{\gamma})$. Therefore, \eqref{equ:013} can be reorganized as
\begin{subequations}\label{equ:015:0}
\begin{align}
\label{equ:015} & \textstyle X = \frac{1}{2} \frac{\partial}{\partial \tau} \left | A_{y s} \right |^2 (\hat{\tau}, \hat{\gamma}) = \text{Re} \left \{ A^*_{s_r s}(\hat{\tau}, \hat{\gamma}) \frac{\partial}{\partial \tau} A_{s_r s}(\hat{\tau}, \hat{\gamma}) \right \}, \\
\label{equ:016} & \textstyle Y = \frac{1}{2} \frac{\partial}{\partial \gamma} \left | A_{y s} \right |^2 (\hat{\tau}, \hat{\gamma}) = \text{Re} \left \{ A^*_{s_r s}(\hat{\tau}, \hat{\gamma}) \frac{\partial}{\partial \gamma} A_{s_r s}(\hat{\tau}, \hat{\gamma}) \right \},
\end{align}
\end{subequations}
where we define
\begin{equation}\label{equ:0065}
\begin{aligned}
& \textstyle X_1 := -\text{Re} \left \{ A^*_{n s}(\hat{\tau}, \hat{\gamma}) \frac{\partial}{\partial \tau} A_{s_r s}(\hat{\tau}, \hat{\gamma}) \right \}, \\
& \textstyle X_2 := -\text{Re} \left \{ A^*_{n s}(\hat{\tau}, \hat{\gamma}) \frac{\partial}{\partial \tau} A_{n s}(\hat{\tau}, \hat{\gamma}) \right \}, \\
& \textstyle X_3 := -\text{Re} \left \{ A^*_{s_r s}(\hat{\tau}, \hat{\gamma}) \frac{\partial}{\partial \tau} A_{n s}(\hat{\tau}, \hat{\gamma}) \right \}, \\
& \textstyle Y_1 := -\text{Re} \left \{ A^*_{n s}(\hat{\tau}, \hat{\gamma}) \frac{\partial}{\partial \gamma} A_{s_r s}(\hat{\tau}, \hat{\gamma}) \right \}, \\
& \textstyle Y_2 := -\text{Re} \left \{ A^*_{n s}(\hat{\tau}, \hat{\gamma}) \frac{\partial}{\partial \gamma} A_{n s}(\hat{\tau}, \hat{\gamma}) \right \}, \\
& \textstyle Y_3 := -\text{Re} \left \{ A^*_{s_r s}(\hat{\tau}, \hat{\gamma}) \frac{\partial}{\partial \gamma} A_{n s}(\hat{\tau}, \hat{\gamma}) \right \}, \\
& \textstyle X := X_1 + X_2 + X_3, \\
& \textstyle Y := Y_1 + Y_2 + Y_3.
\end{aligned}
\end{equation}
As shown by Theorem~\ref{lemma:08}, $\hat{\tau}$ and $\hat{\gamma}$ respectively converge to their true values as $\sigma \rightarrow 0$. This indicates that under high SNRs, $(\hat{\tau}, \hat{\gamma})$ is distributed in a neighborhood of $(\tau_0, \gamma_0)$. Thus we expand the right hand sides (RHSs) of \eqref{equ:015:0} in Taylor series around $(\tau_0, \gamma_0)$, respectively,
\begin{subequations}\label{equ:0064}
\begin{align}
\label{equ:018} & \begin{aligned}
& \textstyle X = \frac{1}{2} \left [ (\hat{\tau} - \tau_0) a(\xi) + (\hat{\gamma} - \gamma_0) b(\xi) \right ], \\
\end{aligned} \\
\label{equ:019} & \begin{aligned}
& \textstyle Y = \frac{1}{2} \left [ (\hat{\tau} - \tau_0) c(\eta) + (\hat{\gamma} - \gamma_0) d(\eta) \right ]. \\
\end{aligned}
\end{align}
\end{subequations}
In \eqref{equ:0064} we respectively define
\begin{equation}
\begin{aligned}
& \textstyle a(\xi) := \frac{\partial^2}{\partial \tau^2} \left | A_{s_r s} \right |^2 (\tau(\xi), \gamma(\xi)), \\
& \textstyle b(\xi) := \frac{\partial^2}{\partial \tau \partial \gamma} \left | A_{s_r s} \right |^2 (\tau(\xi), \gamma(\xi)), \\
& \textstyle c(\eta) := \frac{\partial^2}{\partial \tau \partial \gamma} \left | A_{s_r s} \right |^2 (\tau(\eta), \gamma(\eta)), \\
& \textstyle d(\eta) := \frac{\partial^2}{\partial \gamma^2} \left | A_{s_r s} \right |^2 (\tau(\eta), \gamma(\eta)),
\end{aligned}
\end{equation}
where $\tau(\xi) = \tau_0 + \xi (\hat{\tau} - \tau_0)$, $\gamma(\xi) = \gamma_0 + \xi (\hat{\gamma} - \gamma_0)$, $0 \leq \xi \leq 1$, and $\tau(\eta) = \tau_0 + \eta (\hat{\tau} - \tau_0)$, $\gamma(\eta) = \gamma_0 + \eta (\hat{\gamma} - \gamma_0)$, $0 \leq \eta \leq 1$.
Since $| A_{s_r s}(\tau, \gamma) |^2$ reaches its unique maximum at $(\tau_0, \gamma_0)$ according to Assumption~\ref{assumption:01}, both $\text{Re} \{ A^*_{s_r s}(\tau_0, \gamma_0) \frac{\partial}{\partial \tau} A_{s_r s}(\tau_0, \gamma_0) \}$ and $\text{Re} \{ A^*_{s_r s}(\tau_0, \gamma_0) \frac{\partial}{\partial \gamma} A_{s_r s}(\tau_0, \gamma_0) \}$ equal zero and thus we have directly excluded them from \eqref{equ:0064}.
%
Then we convert \eqref{equ:0064} into the following forms:
\begin{subequations}\label{equ:023:0}
\begin{align}
\label{equ:023} & \textstyle d(\eta) \frac{X}{\sigma} - b(\xi) \frac{Y}{\sigma} = \frac{1}{2} \Lambda(\xi, \eta) \frac{\hat{\tau} - \tau_0}{\sigma}, \\
\label{equ:024} & \textstyle a(\xi) \frac{Y}{\sigma} - c(\eta) \frac{X}{\sigma} = \frac{1}{2} \Lambda(\xi, \eta) \frac{\hat{\gamma} - \gamma_0}{\sigma},
\end{align}
\end{subequations}
where
\begin{equation}\label{equ:0079}
\Lambda(\xi, \eta) := a(\xi) d(\eta) - b(\xi) c(\eta).
\end{equation}
With MSE employed to evaluate the performance of estimation, \eqref{equ:023:0} readily turns into
\begin{subequations}\label{equ:0021:0}
\begin{align}
& \begin{aligned}\label{equ:0021}
& \textstyle \mathbb{E} \left \{ \left | d(\eta) \frac{X}{\sigma} - b(\xi) \frac{Y}{\sigma} \right |^2 \right \} = \textstyle \frac{1}{4} \mathbb{E} \left \{ \left | \Lambda(\xi, \eta) \right |^2 \left | \frac{\hat{\tau} - \tau_0}{\sigma} \right |^2 \right \}, \\
\end{aligned} \\
& \begin{aligned} \label{equ:0022}
& \textstyle \mathbb{E} \left \{ \left | a(\xi) \frac{Y}{\sigma} - c(\eta) \frac{X}{\sigma} \right |^2 \right \} = \textstyle \frac{1}{4} \mathbb{E} \left \{ \left | \Lambda(\xi, \eta) \right |^2 \left | \frac{\hat{\gamma} - \gamma_0}{\sigma} \right |^2 \right \}.
\end{aligned}
\end{align}
\end{subequations}
For each equation in \eqref{equ:0021:0}, we next calculate the limits of both sides as $\sigma \rightarrow 0$, so as to investigate the MSEs of the joint estimation. Based on \eqref{equ:0021}, we focus on the further derivations for the MSE of time delay, while the MSE of Doppler-stretch can be obtained following the similar technique.

\subsection{Calculations for LHS of (\ref{equ:0021})}\label{sec:4.1}

With the definitions in \eqref{equ:0065}, we expand the LHS of (\ref{equ:0021}) into the following form:
\begin{equation}\label{equ:045}
\begin{aligned}
& \textstyle \mathbb{E} \left \{ \left | d(\eta) \frac{X}{\sigma} - b(\xi) \frac{Y}{\sigma} \right |^2 \right \} = \sum_{i, j = 1}^3 \mathbb{E} \left \{ d(\eta) \frac{X_i X_j}{\sigma^2} \right \} + \\
& \quad \textstyle \sum_{i, j = 1}^3 \mathbb{E} \left \{ b(\xi) \frac{Y_i Y_j}{\sigma^2} \right \} - 2 \sum_{i, j = 1}^3 \mathbb{E} \left \{ b(\xi) d(\eta) \frac{X_i Y_j}{\sigma^2} \right \}.
\end{aligned}
\end{equation}
Hence the value of $ \mathbb{E} \{ | d(\eta) \frac{X}{\sigma} - b(\xi) \frac{Y}{\sigma} |^2 \}$ under high SNR is evaluated by successively calculating the limit of each resultant expectation on the RHS of \eqref{equ:045} as $\sigma \rightarrow 0$. Before starting the calculations, it is worth mentioning that since $w(n \Delta)$ follows $\mathbb{C} N (0, \sigma^2)$, the random variable $w(n \Delta) / \sigma$ is thus distributed as $\mathbb{C} N (0, 1)$, which is independent of $\sigma$.



We firstly calculate the limit of $\mathbb{E} \{ d^2(\eta) \frac{X_1^2}{\sigma^2} \}$ as $\sigma \rightarrow 0$. 
For each $\epsilon > 0$, we have
\begin{align}
\nonumber & \textstyle \mathbb{E} \left \{ d^2(\eta) \frac{X_1^2}{\sigma^2} \right \} = \textstyle \mathbb{E} \left \{ d^2(\eta) \frac{X_1^2}{\sigma^2} I_{ \left \{ \left \| (\hat{\tau}, \hat{\gamma}) - (\tau_0, \gamma_0) \right \|_2 > \epsilon \right \} } \right \} \\
\label{equ:0044} & ~~~~~~~~ + \textstyle \mathbb{E} \left \{ d^2(\eta) \frac{X_1^2}{\sigma^2} I_{ \left \{ \left \| (\hat{\tau}, \hat{\gamma}) - (\tau_0, \gamma_0) \right \|_2 \leq \epsilon \right \} } \right \},
\end{align}
where $I_{ \{ \cdot \} }$ denotes the indicator function \cite{Nguyen}. To calculate the first term on the RHS of \eqref{equ:0044}, we know from Theorem~\ref{lemma:08} that
\begin{equation}
\begin{aligned}
& \textstyle \quad d^2(\eta) \frac{X_1^2}{\sigma^2} I_{ \left \{ \left \| (\hat{\tau}, \hat{\gamma}) - (\tau_0, \gamma_0) \right \|_2 > \epsilon \right \} } \stackrel{P}{\longrightarrow} 0 \\
\end{aligned}
\end{equation}
as $\sigma \rightarrow 0$, and
\begin{equation}\label{equ:0011}
\begin{aligned}
& \textstyle \left | d^2(\eta) \frac{X_1^2}{\sigma^2} I_{ \left \{ \left \| (\hat{\tau}, \hat{\gamma}) - (\tau_0, \gamma_0) \right \|_2 > \epsilon \right \} } \right | \\
& \quad \leq \textstyle C_0^2 \frac{X_1^2}{\sigma^2} = C_0^2 \left ( \text{Re} \left \{ A_{\frac{n}{\sigma}, s}^* (\hat{\tau}, \hat{\gamma}) \frac{\partial}{\partial \tau} A_{s_r s}(\hat{\tau}, \hat{\gamma}) \right \} \right )^2 \\
& \quad \leq \textstyle C_0^4 \left ( \max_t \left | s(t) \right | \sum_{n = 0}^{N - 1} \left | \frac{w(n \Delta)}{\sigma} \right | \right )^2,
\end{aligned}
\end{equation}
where

\begin{equation}
\textstyle A_{\frac{n}{\sigma}, s} (\tau, \gamma) := \sum_{n = 0}^{N - 1} \frac{w(n \Delta)}{\sigma} s^*(\gamma (n \Delta - \tau))
\end{equation}
and
\begin{align}
\nonumber C_0 := & \max \left \{ 
\max_{ \tau \in [\tau_{\text{min}}, \tau_{\text{max}}], \gamma \in [\gamma_{\text{min}}, \gamma_{\text{max}}] } \textstyle \left | \frac{\partial}{\partial \tau} A_{s_r s} (\tau, \gamma) \right |, \right . \\
& \left . \max_{ \tau \in [\tau_{\text{min}}, \tau_{\text{max}}], \gamma \in [\gamma_{\text{min}}, \gamma_{\text{max}}] } \textstyle \left | \frac{\partial^2}{\partial \gamma^2} \left | A_{s_r s} \right |^2 (\tau, \gamma) \right | \right \}.
\end{align}
For \eqref{equ:0011}, also note that $\mathbb{E} \{ C_0^4 ( \max_t | s(t)  | \sum_{n = 0}^{N - 1} | \frac{w(n \Delta)}{\sigma} | )^2 \} < + \infty$. Therefore, according to Lebesgue's dominated convergence theorem \cite{Nguyen}, we obtain
\begin{equation}\label{equ:0013}
\lim_{\sigma \rightarrow 0} \textstyle \mathbb{E} \left \{ d^2(\eta) \frac{X_1^2}{\sigma^2} I_{ \left \{ \left \| (\hat{\tau}, \hat{\gamma}) - (\tau_0, \gamma_0) \right \|_2 > \epsilon \right \} } \right \} = 0.
\end{equation}

For the second term on the RHS of \eqref{equ:0044}, on one hand,
\begin{align}\label{equ:0014}
& \liminf_{\sigma \rightarrow 0} \textstyle \mathbb{E} \left \{ d^2(\eta) \frac{X_1^2}{\sigma^2} I_{ \left \{ \left \| (\hat{\tau}, \hat{\gamma}) - (\tau_0, \gamma_0) \right \|_2 \leq \epsilon \right \} } \right \} \\
\nonumber & \quad \geq \min_{\| (\tau, \gamma) - (\tau_0, \gamma_0) \|_2 \leq \epsilon } \textstyle \left ( \frac{\partial^2}{\partial \gamma^2} \left | A_{s_r s}(\tau, \gamma) \right |^2 \right )^2 \displaystyle \liminf_{\sigma \rightarrow 0} \textstyle \mathbb{E} \left \{ \frac{X_1^2}{\sigma^2} \right \}.
\end{align}
Letting $\epsilon \rightarrow 0$ and with \eqref{equ:0013}, we have
\begin{equation}\label{equ:0015_0}
\liminf_{\sigma \rightarrow 0} \textstyle \mathbb{E} \left \{ d^2(\eta) \frac{X_1^2}{\sigma^2} \right \} \geq d^2(0) \displaystyle \liminf_{\sigma \rightarrow 0} \textstyle \mathbb{E} \left \{ \frac{X_1^2}{\sigma^2} \right \}.
\end{equation}
On the other hand, it can be also derived that
%
%
%
%
\begin{equation}\label{equ:0015}
\limsup_{\sigma \rightarrow 0} \textstyle \mathbb{E} \left \{ d^2(\eta) \frac{X_1^2}{\sigma^2} \right \} \leq d^2(0) \displaystyle \limsup_{\sigma \rightarrow 0} \textstyle \mathbb{E} \left \{ \frac{X_1^2}{\sigma^2} \right \}.
\end{equation}
Combining \eqref{equ:0015_0} and \eqref{equ:0015}, we obtain
%
%
\begin{equation}\label{equ:0012}
\begin{aligned}
& d^2(0) \liminf_{\sigma \rightarrow 0} \textstyle \mathbb{E} \left \{ \frac{X_1^2}{\sigma^2} \right \} \leq \displaystyle \liminf_{\sigma \rightarrow 0} \textstyle \mathbb{E} \left \{ d^2(\eta) \frac{X_1^2}{\sigma^2} \right \} \\
& \quad \leq \limsup_{\sigma \rightarrow 0} \textstyle \mathbb{E} \left \{ d^2(\eta) \frac{X_1^2}{\sigma^2} \right \} \leq d^2(0) \displaystyle \limsup_{\sigma \rightarrow 0} \textstyle \mathbb{E} \left \{ \frac{X_1^2}{\sigma^2} \right \}.
\end{aligned}
\end{equation}
It is clear that if $\lim_{\sigma \rightarrow 0} \mathbb{E} \{ \frac{X_1^2}{\sigma^2} \}$ exists, the existence of $\lim_{\sigma \rightarrow 0} \mathbb{E} \{ d^2(\eta) \frac{X_1^2}{\sigma^2} \}$ can be also confirmed. With Theorem~\ref{lemma:08} and \eqref{equ:0011},
\begin{equation}
\begin{aligned}
\textstyle \frac{X_1^2}{\sigma^2} \stackrel{P}{\longrightarrow} \left ( \text{Re} \left \{ A_{\frac{n}{\sigma}, s}^* (\tau_0, \gamma_0) \frac{\partial}{\partial \tau} A_{s_r s}(\tau_0, \gamma_0) \right \} \right )^2
\end{aligned}
\end{equation}
as $\sigma \rightarrow 0$. Again, following Lebesgue's dominated convergence theorem,
\begin{equation}\label{equ:0056}
\lim_{\sigma \rightarrow 0} \textstyle \mathbb{E} \left \{ \frac{X_1^2}{\sigma^2} \right \} = \mathbb{E} \left \{ \left ( \text{Re} \left \{ A_{\frac{n}{\sigma}, s}^* (\tau_0, \gamma_0) \frac{\partial}{\partial \tau} A_{s_r s}(\tau_0, \gamma_0) \right \} \right )^2 \right \}.
\end{equation}
Recall that $w(n \Delta) / \sigma \sim \mathbb{C} N (0, 1)$. As a result, the random variable $A_{\frac{n}{\sigma}, s}^* (\tau_0, \gamma_0) \frac{\partial}{\partial \tau} A_{s_r s}(\tau_0, \gamma_0)$ in \eqref{equ:0056} follows $\mathbb{C}N (0, \sum_{n \in \mathcal{N}}| s(\gamma_0 (n \Delta - \tau_0)) \frac{\partial}{\partial \tau} A_{s_r s}(\tau_0, \gamma_0) |^2)$. Then \eqref{equ:0056} can be calculated as
\begin{align}\label{equ:0047}
\nonumber \lim_{\sigma \rightarrow 0} \textstyle \mathbb{E} \left \{ \frac{X_1^2}{\sigma^2} \right \} = & \textstyle \frac{1}{2} \sum_{n \in \mathcal{N}} \left | s(\gamma_0 (n \Delta - \tau_0)) \frac{\partial}{\partial \tau} A_{s_r s}(\tau_0, \gamma_0) \right |^2 \\
= & \textstyle \frac{1}{2} E_s \left | \frac{\partial}{\partial \tau} A_{s_r s}(\tau_0, \gamma_0) \right |^2,
\end{align}
where we respectively define $E_s := \sum_{n = 0}^{N - 1} | s(\gamma_0 (n \Delta - \tau_0)) |^2$, and $\mathcal{N} := \{n | s(\gamma_0 (n \Delta - \tau_0)) \neq 0, 0 \leq n \leq N - 1, n \in \mathbb{N} \}$. The indices collected by $\mathcal{N}$ in fact corresponds to all the non-zero samples of $s(t)$.
Totally, together with \eqref{equ:0012} and \eqref{equ:0047}, we conclude that $\lim_{\sigma \rightarrow 0} \mathbb{E} \{ d^2(\eta) \frac{X_1^2}{\sigma^2} \}$ exists and
\begin{equation}
\lim_{\sigma \rightarrow 0} \textstyle \mathbb{E} \left \{ d^2(\eta) \frac{X_1^2}{\sigma^2} \right \} = \frac{1}{2} d^2(0) E_s \left | \frac{\partial}{\partial \tau} A_{s_r s}(\tau_0, \gamma_0) \right |^2.
\end{equation}

The calculations for the rest terms in \eqref{equ:045} can be performed similarly. Due to limited space, we eliminate the tedious and repetive calculating processes as presented above, whereas the corresponding results can be found in Appendix~C.

\subsection{Calculations for RHS of (\ref{equ:0021})}\label{sec:4.2}

As for the RHS of \eqref{equ:0021}, given $\epsilon > 0$, we have
\begin{align}\label{equ:075}
& \textstyle \frac{1}{4} \mathbb{E} \left \{ \left | \Lambda(\xi, \eta) \right |^2 \left | \frac{\hat{\tau} - \tau_0}{\sigma} \right |^2 \right \} \\
\nonumber & = \textstyle \frac{1}{4} \mathbb{E} \left \{ \left | \Lambda(\xi, \eta) \right |^2 \left | \frac{\hat{\tau} - \tau_0}{\sigma} \right |^2 I_{ \left \{ \left \| (\hat{\tau}, \hat{\gamma}) - (\tau_0, \gamma_0) \right \|_2 > \epsilon \right \} } \right \} + \\
\nonumber & \quad ~ \textstyle \frac{1}{4} \mathbb{E} \left \{ \left | \Lambda(\xi, \eta) \right |^2 \left | \frac{\hat{\tau} - \tau_0}{\sigma} \right |^2 I_{ \left \{ \left \| (\hat{\tau}, \hat{\gamma}) - (\tau_0, \gamma_0) \right \|_2 \leq \epsilon \right \} } \right \},
\end{align}
%
%
where the two terms on the RHS can be calculated in accordance to the following lemma:
\begin{lem}\label{lemma:001}{\it
For any random variable $H(\omega)$ which is bounded almost everywhere and any $\epsilon > 0$,
\begin{subequations}
\begin{align}
\label{equ:24} & \lim_{\sigma \rightarrow 0} \textstyle \mathbb{E} \left \{ H(\omega) \left | \frac{\hat{\tau} - \tau_0}{\sigma} \right |^2 I_{\{ \omega: \left \| (\hat{\tau}, \hat{\gamma}) - (\tau_0, \gamma_0) \right \|_2 > \epsilon \}} \right \} = 0, \\
& \begin{aligned}\label{equ:22}
& \lim_{\sigma \rightarrow 0} \textstyle \mathbb{E} \left \{ H(\omega) \left | \frac{\hat{\tau} - \tau_0}{\sigma} \right |^2 I_{\{ \omega: \left \| (\hat{\tau}, \hat{\gamma}) - (\tau_0, \gamma_0) \right \|_2 \leq \epsilon \}} \right \} \\
& = \lim_{\sigma \rightarrow 0} \textstyle \mathbb{E} \left \{ H(\omega) \left | \frac{\hat{\tau} - \tau_0}{\sigma} \right |^2 \right \},
\end{aligned}
\end{align}
\end{subequations}
%
%
%
%
where the limit symbols on both sides of \eqref{equ:22} are replaced by superior or inferior limits if the limit on either side of \eqref{equ:22} does not exist.
}\end{lem}
\begin{IEEEproof}
See Appendix~\ref{proof:lemma001}.
\end{IEEEproof}

%
%
By Lemma~\ref{lemma:001}, the first term on the RHS of \eqref{equ:075} directly equals zero.
%
%
%
While for the second term, on one hand,
\begin{equation}\label{equ:082}
\begin{aligned}
& \lim_{\sigma \rightarrow 0} \textstyle \frac{1}{4} \mathbb{E} \left \{ \left | \Lambda(\xi, \eta) \right |^2 \left | \frac{\hat{\tau} - \tau_0}{\sigma} \right |^2 I_{ \left \{ \left \| (\hat{\tau}, \hat{\gamma}) - (\tau_0, \gamma_0) \right \|_2 \leq \epsilon \right \} } \right \} \\
& \geq \frac{1}{4} \min_{ \left \| (\tau_1, \gamma_1) - (\tau_0, \gamma_0) \right \|_2 \leq \epsilon, \left \| (\tau_2, \gamma_2) - (\tau_0, \gamma_0) \right \|_2 \leq \epsilon } \\
& \textstyle \quad \left | \frac{\partial^2}{\partial \tau^2} \left | A_{s_r s} \right |^2 (\tau_1, \gamma_1) \frac{\partial^2}{\partial \gamma^2} \left | A_{s_r s} \right |^2 (\tau_2, \gamma_2) - \right . \\
& \textstyle \quad \left . \frac{\partial^2}{\partial \tau \partial \gamma} \left | A_{s_r s} \right |^2 (\tau_1, \gamma_1) \frac{\partial^2}{\partial \tau \partial \gamma} \left | A_{s_r s} \right |^2 (\tau_2, \gamma_2) \right |^2 \times \\
& \quad \limsup_{\sigma \rightarrow 0} \textstyle \mathbb{E} \left \{ \left | \frac{\hat{\tau} - \tau_0}{\sigma} \right |^2 \right \}.
\end{aligned}
\end{equation}
Since \eqref{equ:082} holds for all $\epsilon > 0$, letting $\epsilon \rightarrow 0$, we have
\begin{equation}\label{equ:0060}
\begin{aligned}
& \lim_{\sigma \rightarrow 0} \textstyle \frac{1}{4} \mathbb{E} \left \{ \left | \Lambda(\xi, \eta) \right |^2 \left | \frac{\hat{\tau} - \tau_0}{\sigma} \right |^2 \right \} \\
& \geq \textstyle \frac{1}{4} \Lambda^2(0, 0) \displaystyle \limsup_{\sigma \rightarrow 0} \textstyle \mathbb{E} \left \{ \left | \frac{\hat{\tau} - \tau_0}{\sigma} \right |^2 \right \}.
\end{aligned}
\end{equation}
On the other hand, we derive
\begin{equation}\label{equ:0061}
\begin{aligned}
& \lim_{\sigma \rightarrow 0} \textstyle \frac{1}{4} \mathbb{E} \left \{ \left | \Lambda(\xi, \eta) \right |^2 \left | \frac{\hat{\tau} - \tau_0}{\sigma} \right |^2 \right \} \\
& \leq \textstyle \frac{1}{4} \Lambda^2(0, 0) \displaystyle \liminf_{\sigma \rightarrow 0} \textstyle \mathbb{E} \left \{ \left | \frac{\hat{\tau} - \tau_0}{\sigma} \right |^2 \right \}.
\end{aligned}
\end{equation}
By \eqref{equ:0060} and \eqref{equ:0061}, it can be concluded that
\begin{equation}\label{equ:089}
\begin{aligned}
\lim_{\sigma \rightarrow 0} & \textstyle \frac{1}{4} \mathbb{E} \left \{ \left | \Lambda(\xi, \eta) \right |^2 \left | \frac{\hat{\tau} - \tau_0}{\sigma} \right |^2 \right \} = \textstyle \frac{1}{4} \Lambda^2(0, 0) \displaystyle \lim_{\sigma \rightarrow 0} \textstyle \mathbb{E} \left \{ \left | \frac{\hat{\tau} - \tau_0}{\sigma} \right |^2 \right \}.
\end{aligned}
\end{equation}

Based on Assumption~\ref{assumption:2.8}, the MSE of time delay can be obtained with the results in Section~\ref{sec:4.1} and \eqref{equ:089}, i.e.,
\begin{align}
& \lim_{\sigma \rightarrow 0} \textstyle \mathbb{E} \left \{ \left | \frac{\hat{\tau} - \tau_0}{\sigma} \right |^2 \right \} \\
\nonumber & = \lim_{\sigma \rightarrow 0} \textstyle \frac{ \mathbb{E} \left \{ d^2(\eta(\omega)) \frac{X^2}{\sigma^2} \right \} + \mathbb{E} \left \{ b^2(\xi(\omega)) \frac{Y^2}{\sigma^2} \right \} - 2 \mathbb{E} \left \{ b(\xi(\omega)) d(\eta(\omega)) \frac{X Y}{\sigma^2} \right \} }{ \textstyle \frac{1}{4} \Lambda^2(0, 0) }.
\end{align}
Lastly, letting the sampling rate become sufficiently large and employing the definitions in \eqref{equ:0066}, we eventually derive the integration representation of the MSE in \eqref{equ:0034a}.

\subsection{Approximation in (\ref{equ:0034b})}\label{proof:theorem02}

We next show that the approximation conducted in \eqref{equ:0034b} is a proper one. 
Based on Assumption~\ref{assumption:04}, we substitute \eqref{equ:16} into $D$ and $E$, respectively, resulting in
%
%
%
\begin{subequations}
\begin{align}
& \textstyle D = \sum_{k = 0}^{K - 1} \int_{0}^{T} (t + k T_r)^2 \left [ \dot{\beta}^2 (t) + 4 \pi^2 f_k^2 \beta^2(t) \right ] dt, \\
&\textstyle E = K \int_{0}^{T} \beta^2(t) dt.
\end{align}
\end{subequations}
Assuming that the envelope $\beta(t)$ is a symmetric function, i.e. $\beta(t) = \beta(T - t)$, we perform the derivation as follows:
\begin{equation}\label{equ:0062}
\begin{aligned}
\frac{D}{E} & \geq \frac{ 4 \pi^2 \sum_{k = 0}^{K - 1} f_k^2 \int_{0}^{T} (t + k T_r)^2 \beta^2(t) dt }{ K \int_{0}^{T} \beta^2(t) dt } \\
& \geq 4 \pi^2 \frac{1}{K} \sum_{k = 0}^{K - 1} f_k^2 \frac{ \int_{0}^{T} t^2 \beta^2(t) dt }{ \int_{0}^{T} \beta^2(t) dt } \\
& \approx 4 \pi^2 f_0^2 \frac{ \int_{0}^{T/2} t^2 \beta^2(t) dt + \int_{0}^{T/2} (T - t)^2 \beta^2(t) dt }{ 2 \int_{0}^{T/2} \beta^2(t) dt } \\
& \geq 4 \pi^2 f_0^2 \frac{ \frac{T^2}{2} \int_{0}^{T/2} \beta^2(t) dt }{ 2 \int_{0}^{T/2} \beta^2(t) dt } = \pi^2 T^2 f_0^2 \gg 1.
\end{aligned}
\end{equation}
The last line of \eqref{equ:0062} is generally satisfied for a radar signal, since the central carrier frequency $f_0$ is usually several orders of magnitude larger than $1 / T$.
Therefore, with the fact that $D \gg E$, the terms with $E^2$ in \eqref{equ:0034a} are much smaller compared to those with $E D$ and thus can be eliminated. The approximation made in \eqref{equ:0034b} is hence a proper one, and the performance of the AF-based estimation can be evaluated by the much simplified forms of MSEs as given by \eqref{equ:0034b}.

\subsection{Proof of Theorem~\ref{theorem:03}}\label{proof:theorem03}

We utilize the symmetry property of the envelope $\beta(t)$. As $\beta(t) = \beta(T - t)$ is satisfied in most cases, we can also derive that $\dot{\beta}(t) = -\dot{\beta}(T - t)$. As a result, the following relations are readily satisfied:
\begin{subequations}\label{equ:0066}
\begin{align}
& \textstyle \int_{0}^{T} \left (t - \frac{T}{2} \right ) \beta^2(t) dt = 0, \\
& \textstyle \int_{0}^{T} \left (t - \frac{T}{2} \right ) \dot{\beta}^2(t) dt = 0.
\end{align}
\end{subequations}
With \eqref{equ:0066}, we respectively obtain
\begin{subequations}\label{equ:0067}
\begin{equation}
\label{equ:0067:1}
\begin{aligned}
& \textstyle B - \frac{F^2}{E} = K \left [ S_0^{(1)} + 4 \pi^2 \text{Var}\left \{ f_k \right \} S_0^{(0)} \right ],
\end{aligned}
\end{equation}
\begin{align}
\nonumber & \textstyle D - \frac{G^2}{E} = \textstyle K S_2^{(1)} + 4 \pi^2 \sum_{k = 0}^{K - 1} f_k^2 \int_{0}^{T} \left (t - \frac{T}{2} \right )^2 \beta^2(t) dt + \\
\nonumber & \textstyle \quad 4 \pi^2 K \Omega_1 S_0^{(0)} + K \left [ \left ( \frac{1}{K} \sum_{k = 0}^{K - 1} T_k \right )^2 S_0^{(1)} + \right . \\
\label{equ:0067:3} & \textstyle \quad \left . 4 \pi^2 \frac{ 1 }{ \text{Var} \left \{ f_k \right \} } \text{Cov} \left \{ f_k, f_k T_k \right \} S_0^{(0)} \right ],
\end{align}
\begin{align}
\nonumber & \textstyle \frac{\Pi}{E^2} = \textstyle K \left (B - \frac{F^2}{E} \right ) \left [ S_2^{(1)} + 4 \pi^2 \Omega_1 S_0^{(0)} + \right . \\
\label{equ:0069} & \textstyle \left . 4 \pi^2 \frac{1}{K} \sum_{k = 0}^{K - 1} f_k^2 \int_{0}^{T} \left (t - \frac{T}{2} \right )^2 \beta^2(t) dt \right ] + 4 \pi^2 K^2 \Omega_2^2 S_0^{(1)} S_0^{(0)},
\end{align}
\end{subequations}
where 
\begin{subequations}
\begin{align}
& \textstyle \Omega_1 = \left ( 1 - \rho^2(f_k, f_k T_k) \right ) \text{Var}\left \{ f_k T_k \right \}, \\
& \textstyle \Omega_2 = \text{Std}\left \{ f_k \right \} \frac{1}{K} \sum_{k = 0}^{K - 1} T_k - \rho(f_k, f_k T_k) \text{Std}\left \{ f_k T_k \right \}.
\end{align}
\end{subequations}
For \eqref{equ:0067:3} and \eqref{equ:0069}, further simplifications can be conducted. 
According to Assumption~\ref{assumption:2.7}, we derive that
\begin{equation}\label{equ:0070}
\text{Cov}\left \{ f_k, T_k \right \} = \text{Cov}\left \{ f_k^2, T_k \right \} = \text{Cov}\left \{ f_k^2, T_k^2 \right \} = 0.
\end{equation}
As a result,
\begin{align}\label{equ:0071}
\nonumber & \Omega_2 = \textstyle \frac{1}{ \text{Std}\left \{ f_k \right \} } \left [ \text{Var}\left \{ f_k \right \} \frac{1}{K} \sum_{k = 0}^{K - 1} T_k - \text{Cov}\left \{ f_k, f_k T_k \right \} \right ] \\
& = \textstyle \frac{1}{ \text{Std}\left \{ f_k \right \} } \left [ \text{Cov}\left \{ f_k, T_k \right \} \frac{1}{K} \sum_{k = 0}^{K - 1} f_k - \text{Cov}\left \{ f_k^2, T_k \right \} \right ] = 0.
\end{align}
The second term on the RHS of \eqref{equ:0069} thus equals zero. On the other hand, since $\text{Var}\left \{ f_k \right \} \frac{1}{K} \sum_{k = 0}^{K - 1} T_k = \text{Cov}\left \{ f_k, f_k T_k \right \}$ (as revealed by the first line of \eqref{equ:0071}), and together with \eqref{equ:0070},
%
%
\begin{align}\label{equ:0072}
\nonumber & \Omega_1 = \textstyle \text{Var}\left \{ f_k T_k \right \} - \text{Var}\left \{ f_k \right \} \left ( \frac{1}{K} \sum_{k = 0}^{K - 1} T_k \right )^2 \\
\nonumber & = \textstyle \text{Cov}\left \{ f_k^2, T_k^2 \right \} + \frac{1}{K} \sum_{k = 0}^{K - 1} f_k^2 \text{Var}\left \{ T_k \right \} - \text{Cov}\left \{ f_k, T_k \right \} \times \\
\nonumber & \quad \textstyle \left [ \left ( \frac{1}{K} \sum_{k = 0}^{K - 1} f_k \right ) \left ( \frac{1}{K} \sum_{k = 0}^{K - 1} T_k \right ) + \frac{1}{K} \sum_{k = 0}^{K - 1} f_k T_k \right ] \\
& = \textstyle \frac{1}{K} \sum_{k = 0}^{K - 1} f_k^2 \text{Var}\left \{ T_k \right \}.
\end{align}
Substituting \eqref{equ:0071}--\eqref{equ:0072} into \eqref{equ:0067:3} and \eqref{equ:0069}, we eventually have
\begin{subequations}\label{equ:0073}
\begin{align}
\nonumber & \textstyle D - \frac{G^2}{E} = \textstyle K \left [ S_2^{(1)} + 4 \pi^2 \frac{1}{K} \sum_{k = 0}^{K - 1} f_k^2 S_2^{(0)} \right ] + \\
& ~~~~~~ \textstyle \left ( \frac{1}{K} \sum_{k = 0}^{K - 1} T_k \right )^2 \left ( B - \frac{F^2}{E} \right ), \\
& \textstyle \frac{\Pi}{E^2} = \textstyle K \left ( B - \frac{F^2}{E} \right ) \left [ S_2^{(1)} + 4 \pi^2 \frac{1}{K} \sum_{k = 0}^{K - 1} f_k^2 S_2^{(0)} \right ].
\end{align}
\end{subequations}
%
%
%
%
%
%
Lastly, by substituting \eqref{equ:0067:1} and \eqref{equ:0073} into \eqref{equ:0034b}, the compact expressions of MSEs as presented by Theorem~\ref{theorem:03} are concluded.

\section{Numerical Results}

In this section, we verify the correctness of our main results by numerical examples. Set the envelope of the RSF signal $\beta(t)$ as
\begin{equation}
\beta(t) = \left \{
\begin{array}{ll}
t^3 (T - t)^3, & \text{if } t \in [0, T], \\
0, & \text{otherwise},
\end{array}
\right .
\end{equation}
which not only attains a rectangle-like shape, but also satisfies all the assumptions in Section~II-B. The SNR at the receiver is defined as
\begin{equation}
\text{SNR} := \frac{1}{N_0} \int_{-\infty}^{\infty} \left | x s(\gamma_0 (t - \tau_0)) \right |^2 dt
\end{equation}
throughout the simulations \cite{Luo95}. Some main parameters of the simulation environment are configured in TABLE~\ref{tabel:01}.

\begin{figure}[t]
  \centering
  \includegraphics[width=2.8 in]{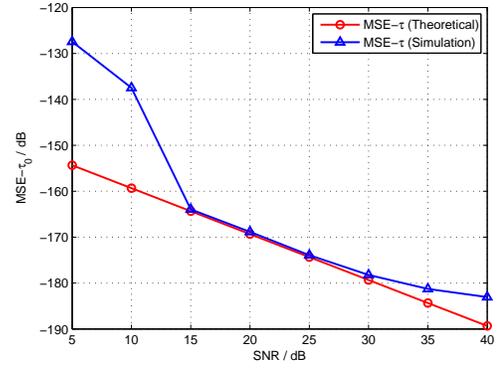}\\
  \caption{Theoretical and simulated MSEs of time delay.}\label{tau_simu}
\end{figure}

\begin{figure}[t]
  \centering
  \includegraphics[width=2.8 in]{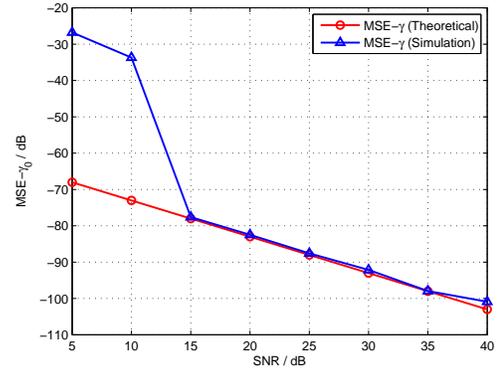}\\
  \caption{Theoretical and simulated MSEs of Doppler-stretch.}\label{gamma_simu}
\end{figure}

We first compare the MSEs of the AF-based estimation \eqref{equ:0063} with their theoretical counterparts (as given by Theorem~\ref{theorem:02} or \ref{theorem:03}) by simulations. Under each of the SNRs from $5$ dB to $40$ dB, the simulated and theoretical MSEs are calculated on the basis of $200$ independent Monte Carlo simulations. In each trial, a Costas frequency shifting codeword is randomly generated from a set $\mathcal{C} = \{ -2.5, -1.5, -0.5, 0.5, 1.5, 2.5 \}$ \cite{Golomb84}. The central carrier frequency $f_0$ and the minimum frequency shifting step $\delta_f$ are respectively set to $20$ MHz and $2$ MHz. As depicted in Figs.~\ref{tau_simu} and \ref{gamma_simu}, when $\text{SNR} \geq 15$ dB, both the simulated MSEs of time delay and Doppler-stretch perfectly converge to their theoretical counterparts, indicating that the performance of the AF-based estimation with the RSF signal under high SNRs can be accurately described by Theorems~\ref{theorem:02}--\ref{theorem:03}.


\begin{table}[b]
\caption{Parameter settings for simulations.}\label{tabel:01}
\small
\centering
\begin{tabular}[c]{p{0.25 \textwidth} | p{0.15 \textwidth}}
\hline
Parameters & Values \\
\hline
Time delay $\tau_0$ & $1.48 \times 10^{-6}$ sec \\
Doppler-stretch $\gamma_0$ & $0.91$ \\
Duration of envelope $T$ & $1 \times 10^{-6}$ sec \\
PRI $T_r$ & $4 \times 10^{-6}$ sec \\
Scattering coefficient $x$ & $1$ \\
Number of pulses $K$ & $6$ \\
Sampling interval $\Delta$ & $1 \times 10^{-8}$ sec \\
\hline
\end{tabular}
\vspace{0.2cm}
\end{table}

\subsection{Influence of Waveform Parameters}

Next, we investigate the influence of the waveform parameters on the estimation performance with the RSF signal.

\begin{figure}[t]
  \centering
  \includegraphics[width=2.8 in]{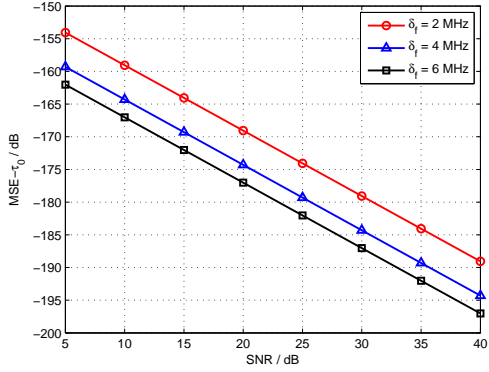}\\
  \caption{MSEs of time delay with different bandwidths.}\label{tau_bandwidth}
\end{figure}

\begin{figure}[t]
  \centering
  \includegraphics[width=2.8 in]{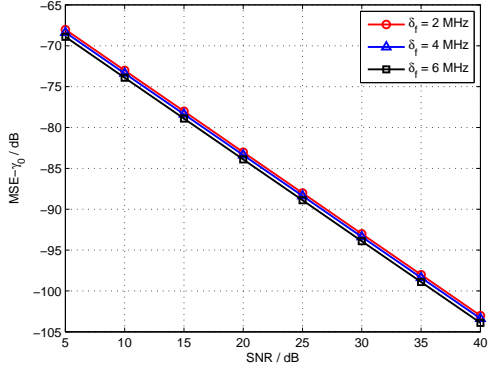}\\
  \caption{MSEs of Doppler-stretch with different bandwidths.}\label{gamma_bandwidth}
\end{figure}

In order to reveal the relation between the estimation performance and the bandwidth of the carrier frequencies, we fix $f_0$ to $20$ MHz and set $\delta_f$ to $[2, 4, 6]$ MHz, respectively, in three sets of simulations. The Costas codeword is still adopted. Figs.~\ref{tau_bandwidth} and \ref{gamma_bandwidth} illustrate the estimation performances with the three different bandwidths. Just as predicted by Theorem~\ref{theorem:03}, under each SNR value, with the increasing of $\delta_f$, the MSE of time delay decreases, while there is no significant variation found in the MSEs of Doppler-stretch.

\begin{figure}[t]
  \centering
  \includegraphics[width=2.8 in]{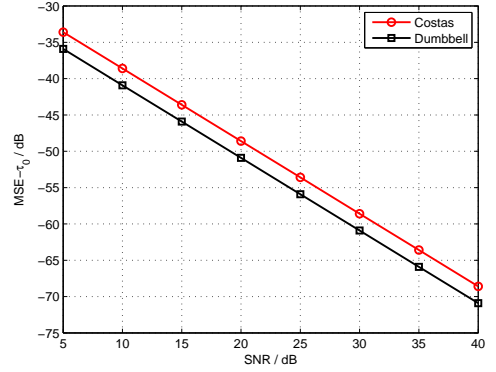}\\
  \caption{MSEs of time delay with different codewords.}\label{tau_codeword}
\end{figure}

\begin{figure}[t]
  \centering
  \includegraphics[width=2.8 in]{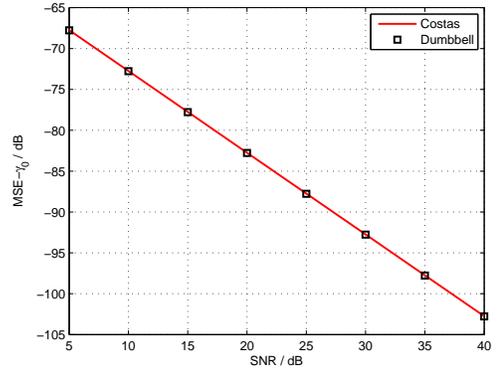}\\
  \caption{MSEs of Doppler-stretch with different codewords.}\label{gamma_codeword}
\end{figure}

Within a fixed bandwidth covered by the carrier frequencies, we explore the impact of different frequency shifting patterns on the estimation performance. Besides the Costas codeword as generated above, we also involve a Dumbbell codeword for comparison, with which only the lowest and the highest frequencies (i.e. $f_0 - 2.5 \delta_f$ and $f_0 + 2.5 \delta_f$) are available for using. Set $f_0$ and $\delta_f$ to $20$ MHz and $2$ MHz, respectively. The estimation performances with the two frequency shifting codewords are shown in Figs.~\ref{tau_codeword} and \ref{gamma_codeword}. The Dumbell codeword always outperforms the Costas codeword in delay estimating, while the performances of Doppler-stretch estimation with the two codewords are almost the same. This is consistent with our conclusion that a randomized codeword with larger variance leads to a better performance of delay estimation.

\begin{figure}[t]
  \centering
  \includegraphics[width=2.8 in]{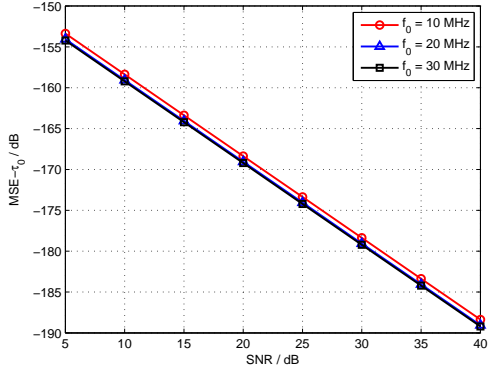}\\
  \caption{MSEs of time delay with different central carrier frequencies.}\label{tau_carrier}
\end{figure}

\begin{figure}[t]
  \centering
  \includegraphics[width=2.8 in]{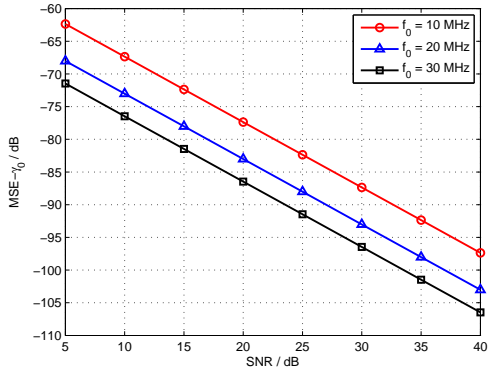}\\
  \caption{MSEs of Doppler-stretch with different central carrier frequencies.}\label{gamma_carrier}
\end{figure}

Then we compare the estimation performances with different central carrier frequencies. Set $f_0$ to $[10, 20, 30]$ MHz, respectively, while fix the minimum frequency shifting step $\delta_f$ to $2$ MHz. The estimation performances with the Costas encoded RSF signal are shown in Figs.~\ref{tau_carrier} and \ref{gamma_carrier}. As $f_0$ increases, the MSE of Doppler-stretch evidently descends, while that of time delay only slightly decreases. Beyond the revealings by Theorem~\ref{theorem:03}, we summarise from the simulation results that the central carrier frequency mainly determines the performance of Doppler-stretch estimation, while its impact on the performance of delay estimation is negligible. Instead, it is the bandwidth covered by the carrier frequencies that dominates the performance of delay estimation.

\subsection{Comparison with Other Waveforms}

\begin{figure}[t]
  \centering
  \includegraphics[width=2.8 in]{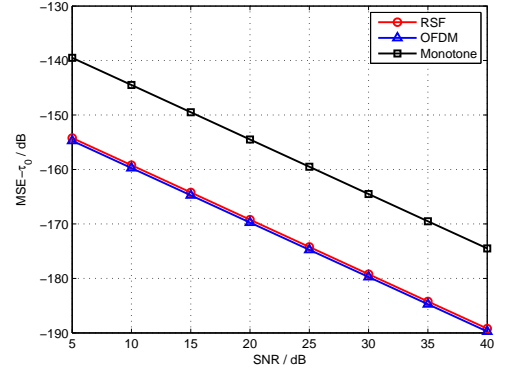}\\
  \caption{MSEs of time delay with different waveforms.}\label{tau_waveform}
\end{figure}

\begin{figure}[t]
  \centering
  \includegraphics[width=2.8 in]{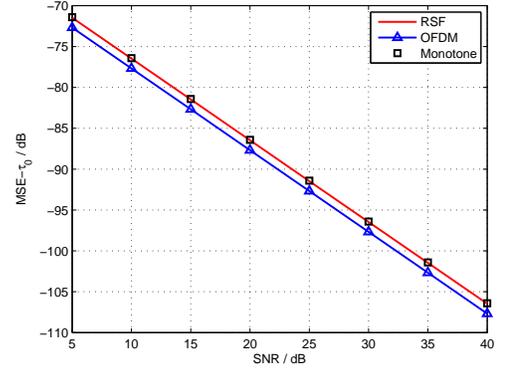}\\
  \caption{MSEs of Doppler-stretch with different waveforms.}\label{gamma_waveform}
\end{figure}

Figs.~\ref{tau_waveform} and \ref{gamma_waveform} compare the estimation performances with the RSF, OFDM and monotone signals. For the RSF signal, reset $f_0$ and $\delta_f$ to $20$ MHz and $2$ MHz, respectively, and still adopt the Costas codeword. The subcarriers of the OFDM signal satisfy $\{ d_l; l = 0, 1, \ldots, L - 1 \} = \mathcal{C}$, while the carrier frequency of the monotone signal is $20$ MHz. In terms of delay estimation as shown in Fig.~\ref{tau_waveform}, due to the wideband character, the RSF signal and the OFDM signal both outperform the narrowband monotone signal, leading by up to $15$ dB of performance gain. It is also observed that the performance with the RSF signal is almost the same as that with the OFDM signal, since the two signals take up the same bandwidth of the carrier frequencies. For the Doppler-stretch estimation as illustrated by Fig.~\ref{gamma_waveform}, the three signals achieve the similar estimation performances. This can be explained by \eqref{equ:theorem3:2}, \eqref{equ:monotone:2}, and \eqref{equ:mse_ofdm:2}, where the identical central carrier frequencies of the three signals contribute to the similar performances of Doppler-stretch estimation.

\section{Concluding Remarks}

We investigated the performance of joint delay-Doppler estimation with the RSF signals. Compact expressions of MSEs with respect to time delay and Doppler-stretch were obtained, revealing the major waveform parameters that influence the estimation performance. Since the derived theoretical MSEs were consistent with the corresponding CRLBs, the AF-based estimation was shown approximately efficient. So as to improve the estimation performance with the RSF signal, higher carrier frequencies, wider bandwidth covered by the carrier frequencies, as well as codewords with larger variance are expected. The RSF signal achieves the same estimation performance as the OFDM signal with only a narrow instantaneous bandwidth. Requirement for the bandwidth of the receiver is thus largely reduced in practical radar systems.

\appendix

\subsection{Proof of Theorem~\ref{lemma:08}}\label{proof:theorem01}

We first investigate the gap between $\left | A_{y s}(\hat{\tau}, \hat{\gamma}) \right |$ and $\left | A_{s_r s}(\tau_0, \gamma_0) \right |$ in the $L^p$ sense.
%
%
%
%
\begin{equation}
\begin{aligned}
& \max_{ \tau \in [\tau_{\text{min}}, \tau_{\text{max}}], \gamma \in [\gamma_{\text{min}}, \gamma_{\text{max}}] } \left | A_{y s}(\tau, \gamma) \right | \geq \left | A_{y s}(\tau_0, \gamma_0) \right | \\
& \geq \textstyle \left | A_{s_r s}(\tau_0, \gamma_0) \right | - \left | A_{n s}(\tau_0, \gamma_0) \right | \\
& = \textstyle \left | A_{s_r s}(\tau_0, \gamma_0) \right | - \left | \sum_{n = 0} ^ {N - 1} w(n \Delta) s^*(\gamma_0 (n \Delta - \tau_0)) \right |. \\
\end{aligned}
\end{equation}
%
%
%
%
\begin{equation}
\begin{aligned}
& \textstyle \left | A_{y s}(\hat{\tau}, \hat{\gamma}) \right | \\
= & \textstyle \left | A_{s_r s}(\hat{\tau}, \hat{\gamma}) + \sum_{n = 0} ^ {N - 1} w(n \Delta) s^*(\hat{\gamma} (n \Delta - \hat{\tau})) \right | \\
\leq & \textstyle \left | A_{s_r s}(\tau_0, \gamma_0) \right | + \left | \sum_{n = 0} ^ {N - 1} w(n \Delta) s^*(\hat{\gamma} (n \Delta - \hat{\tau})) \right |. \\
\end{aligned}
\end{equation}
Thus,
\begin{equation}
\begin{aligned}
& \textstyle \left | \left | A_{y s}(\hat{\tau}, \hat{\gamma}) \right | - \left | A_{s_r s}(\tau_0, \gamma_0) \right | \right | \\
& \leq \textstyle \left | \sum_{n = 0} ^ {N - 1} w(n \Delta) s^*(\hat{\gamma} (n \Delta - \hat{\tau})) \right |.
\end{aligned}
\end{equation}
This also implies that
\begin{equation}
\begin{aligned}
& \textstyle \mathbb{E} \left \{ \left | \left | A_{y s}(\hat{\tau}, \hat{\gamma}) \right | - \left | A_{s_r s}(\tau_0, \gamma_0) \right | \right |^p \right \} \\
& \leq \textstyle \mathbb{E} \left \{ \left ( \sum_{n = 0} ^ {N - 1} \left | w(n \Delta) s^*(\hat{\gamma} (n \Delta - \hat{\tau})) \right | \right )^p \right \} \\
& \leq \textstyle N^{p - 1} \left ( \max_t \left | s(t) \right | \right )^p \mathbb{E} \left \{ \sum_{n = 0} ^ {N - 1} \left | w(n \Delta) \right |^p \right \}. \\
\end{aligned}
\end{equation}
Recall that $w(n \Delta) \sim \mathbb{C} N (0, \sigma^2)$. The PDF of $\left | w(n \Delta) \right |$ is then
\begin{equation}
p(\eta) = \frac{2 \eta}{\sigma^2} e^{-\frac{\eta^2}{\sigma^2}}, \quad \eta > 0.
\end{equation}
As a result,
\begin{equation}
\textstyle \mathbb{E} \left \{ \left | w(n \Delta) \right |^p \right \} = \sigma^p \Gamma(\frac{p}{2} + 1).
\end{equation}
For $p \geq 1$,
\begin{equation}\label{lemma:11}
\begin{aligned}
& \textstyle \mathbb{E} \left \{ \left | \left | A_{y s}(\hat{\tau}, \hat{\gamma}) \right | - \left | A_{s_r s}(\tau_0, \gamma_0) \right | \right |^p \right \} \\
& \leq \textstyle N^{p} \sigma^p \Gamma(\frac{p}{2} + 1) \left ( \max_t \left | s(t) \right | \right )^p.
\end{aligned}
\end{equation}

Next, we investigate the gap between $| A_{s_r s}(\hat{\tau}, \hat{\gamma})|$ and $|A_{s_r s}(\tau_0, \gamma_0)|$.

\begin{equation}
\begin{aligned}
& \mathbb{E} \left \{ \left | | A_{s_r s}(\hat{\tau}, \hat{\gamma})| - |A_{s_r s}(\tau_0, \gamma_0)| \right |^p \right \} \\
& \leq \mathbb{E} \left \{ \left ( \left | |A_{s_r s}(\hat{\tau}, \hat{\gamma})| - |A_{y s}(\hat{\tau}, \hat{\gamma})| \right | + \right . \right . \\
& \quad \left . \left . \left | |A_{y s}(\hat{\tau}, \hat{\gamma})| - |A_{s_r s}(\tau_0, \gamma_0)| \right | \right )^p \right \} \\
& \leq 2^{p - 1} \left ( \mathbb{E} \left \{ \left | |A_{s_r s}(\hat{\tau}, \hat{\gamma})| - |A_{y s}(\hat{\tau}, \hat{\gamma})| \right |^p \right \} + \right . \\
& \quad \left . \mathbb{E} \left \{ \left | |A_{y s}(\hat{\tau}, \hat{\gamma})| - |A_{s_r s}(\tau_0, \gamma_0)| \right |^p \right \} \right ),
\end{aligned}
\end{equation}
where
\begin{equation}
\begin{aligned}
& \mathbb{E} \left \{ \left | |A_{s_r s}(\hat{\tau}, \hat{\gamma})| - |A_{y s}(\hat{\tau}, \hat{\gamma})| \right |^p \right \} \leq \mathbb{E} \left \{ \left | A_{n s}(\hat{\tau}, \hat{\gamma}) \right |^p \right \} \\
& \leq \textstyle \left ( \max_t \left | s(t) \right | \right )^p \mathbb{E} \left \{ \left ( \sum_{n = 0}^{N - 1} \left | w(n \Delta) \right | \right )^p \right \} \\
& \leq \textstyle N^{p - 1} \left ( \max_t \left | s(t) \right | \right )^p \mathbb{E} \left \{ \sum_{n = 0} ^ {N - 1} \left | w(n \Delta) \right |^p \right \} \\
& = \textstyle N^{p} \sigma^p \Gamma(\frac{p}{2} + 1) \left ( \max_t \left | s(t) \right | \right )^p.
\end{aligned}
\end{equation}
Thus, we conclude that for $p \geq 1$,
\begin{equation}\label{equ:18}
\begin{aligned}
& \textstyle \mathbb{E} \left \{ \left | | A_{s_r s}(\hat{\tau}, \hat{\gamma})| - |A_{s_r s}(\tau_0, \gamma_0)| \right |^p \right \} \\
& \leq \textstyle 2^p N^{p} \sigma^p \Gamma(\frac{p}{2} + 1) \left ( \max_t \left | s(t) \right | \right )^p.
\end{aligned}
\end{equation}

Then we prove Theorem~\ref{lemma:08} by contradiction. Suppose $(\hat{\tau}, \hat{\gamma})$ does not converge to $(\tau_0, \gamma_0)$ in probability as $\sigma \rightarrow 0$, then there exists $\epsilon_1 > 0$ such that
\begin{equation}
P(\|(\hat{\tau}, \hat{\gamma}) - (\tau_0, \gamma_0) \|_2 > \epsilon_1) \nrightarrow 0
\end{equation}
as $\sigma \rightarrow 0$. Therefore, there exist an $\epsilon_2 > 0$ and a sequence $\left \{ \sigma_n \right \}_{n \geq 1}$ satisfying $\sigma_n \rightarrow 0$ as $n \rightarrow +\infty$, such that
\begin{equation}
P(\|(\hat{\tau}, \hat{\gamma}) - (\tau_0, \gamma_0) \|_2 > \epsilon_1) \geq \epsilon_2
\end{equation}
holds for all $n \geq 1$. Then, for the aforementioned $\epsilon_1$, there exists $C(\epsilon_1) > 0$, such that for all $(\tau, \gamma)$ satisfying $\|(\tau, \gamma) - (\tau_0, \gamma_0) \|_2 \geq \epsilon_1$,
\begin{equation}
\left | | A_{s_r s}(\tau, \gamma)| - |A_{s_r s}(\tau_0, \gamma_0)| \right | \geq C(\epsilon_1).
\end{equation}
Thus,
\begin{align}
\nonumber & \mathbb{E} \left \{ \left | | A_{s_r s}(\hat{\tau}, \hat{\gamma})| - |A_{s_r s}(\tau_0, \gamma_0)| \right |^p \right \} \\
\nonumber & \geq \mathbb{E} \left \{ \left | | A_{s_r s}(\hat{\tau}, \hat{\gamma})| - |A_{s_r s}(\tau_0, \gamma_0)| \right |^p I_{\{ \left | (\hat{\tau}, \hat{\gamma}) - (\tau_0, \gamma_0) \right | > \epsilon_1 \}} \right \} \\
& \geq \left [ C(\epsilon_1) \right ]^p \epsilon_2,
\end{align}
implying that $\mathbb{E} \left \{ \left | | A_{s_r s}(\hat{\tau}, \hat{\gamma})| - |A_{s_r s}(\tau_0, \gamma_0)| \right |^p \right \} \nrightarrow 0$ as $n \rightarrow +\infty$, which contradicts \eqref{equ:18}. Theorem~\ref{lemma:08} is thus proved.

\subsection{Proof of Lemma~\ref{lemma:001}}\label{proof:lemma001}

\begin{align}
\nonumber & \textstyle \mathbb{E} \left \{ \left | H(\omega) \right | \left | \frac{\hat{\tau} - \tau_0}{\sigma} \right |^2 I_{\{ \omega: \left \| (\hat{\tau}, \hat{\gamma}) - (\tau_0, \gamma_0) \right \|_2 > \epsilon \}} \right \} \\
\nonumber & \leq \textstyle \mathbb{E} \left \{ \left | H(\omega) \right | \left | \frac{\hat{\tau} - \tau_0}{\sigma} \right |^2 I_{\{ \omega: \left | | A_{s_r s}(\hat{\tau}, \hat{\gamma})| - |A_{s_r s}(\tau_0, \gamma_0)| \right | > C_{\epsilon} \}} \right \} \\
& \leq \textstyle \left ( \mathbb{E} \left \{ \left | H(\omega) \right |^2 \left | \hat{\tau} - \tau_0 \right |^4 \right \} \right )^{\frac{1}{2}} \times \\
\nonumber & \quad \textstyle \frac{1}{\sigma^2} \left [ P \left ( \omega: \left | | A_{s_r s}(\hat{\tau}, \hat{\gamma})| - |A_{s_r s}(\tau_0, \gamma_0)| \right | > C_{\epsilon} \right ) \right ]^{\frac{1}{2}}.
\end{align}
For $p > 4$, according to \eqref{lemma:11}, we have
\begin{align}
\nonumber & \lim_{\sigma \rightarrow 0} \textstyle \frac{1}{\sigma^2} \left [ P \left ( \omega: \left | | A_{s_r s}(\hat{\tau}, \hat{\gamma})| - |A_{s_r s}(\tau_0, \gamma_0)| \right | > C_{\epsilon} \right ) \right ]^{\frac{1}{2}} \\
& \leq \lim_{\sigma \rightarrow 0} \textstyle \frac{1}{\sigma^2} \left [ \mathbb{E} \left \{ \left | | A_{s_r s}(\hat{\tau}, \hat{\gamma})| - |A_{s_r s}(\tau_0, \gamma_0)| \right |^p \right \} C_{\epsilon}^{-p} \right ]^{\frac{1}{2}} \\
\nonumber & \leq \lim_{\sigma \rightarrow 0} \textstyle \sigma^{\frac{p}{2} - 2} \left [ 2^p N^{p} \Gamma(\frac{p}{2} + 1) \left ( \max_t \left | s(t) \right | \right )^p C_{\epsilon}^{-p} \right ]^{\frac{1}{2}} = 0.
\end{align}
Therefore, \eqref{equ:24} and \eqref{equ:22} are concluded. 

\subsection{Calculations for (\ref{equ:045})}\label{appendix:calculation}

Imitating the calculation for $\mathbb{E} \{ d^2(\eta(\omega)) \frac{X_1^2}{\sigma^2} \}$ in Section~IV-A, we obtain the following results:
\begin{subequations}
\begin{equation}
\begin{aligned}
& \lim_{\sigma \rightarrow 0} \textstyle \mathbb{E} \left \{ d^2(\eta(\omega)) \frac{X_3^2}{\sigma^2} \right \} \\
& = \textstyle d^2(0) \frac{E_s^2 \gamma_0^2 \left | x \right |^2}{2} \sum_{n \in \mathcal{N}} \left | \dot{s}(\gamma_0 (n \Delta - \tau_0)) \right |^2, \\
\end{aligned}
\end{equation}
\begin{equation}
\lim_{\sigma \rightarrow 0} \textstyle \mathbb{E} \left \{ b^2(\xi(\omega)) \frac{Y_1^2}{\sigma^2} \right \} = b^2(0) \frac{E_s}{2} \left | \frac{\partial}{\partial \gamma} A_{s_r s}(\tau_0, \gamma_0) \right |^2,
\end{equation}
\begin{equation}
\begin{aligned}
& \lim_{\sigma \rightarrow 0} \textstyle \mathbb{E} \left \{ b^2(\xi(\omega)) \frac{Y_3^2}{\sigma^2} \right \} \\
& = \textstyle b^2(0) \frac{E_s^2 |x|^2}{2} \sum_{n \in \mathcal{N}} \left | (n \Delta - \tau_0) \dot{s}(\gamma_0 (n \Delta - \tau_0)) \right |^2,
\end{aligned}
\end{equation}
\begin{equation}\label{equ:0048}
\begin{aligned}
& \lim_{\sigma \rightarrow 0} \textstyle \mathbb{E} \left \{ d^2(\eta(\omega)) \frac{X_1 X_3}{\sigma^2} \right \} = d^2(0) \frac{\gamma_0^2 \left | x \right |^2 E_s}{2} \times \\
& \textstyle \text{Re} \left \{ \left ( \sum_{n \in \mathcal{N}} s(\gamma_0 (n \Delta - \tau_0)) \dot{s}^*(\gamma_0 (n \Delta - \tau_0)) \right )^2 \right \}, \\
\end{aligned}
\end{equation}
\begin{align}
& \lim_{\sigma \rightarrow 0} \textstyle \mathbb{E} \left \{ b^2(\xi(\omega)) \frac{Y_1 Y_3}{\sigma^2} \right \} = b^2(0) \frac{\left | x \right |^2 E_s}{2} \times \\
\nonumber & \textstyle \text{Re} \left \{ \left ( \sum_{n \in \mathcal{N}} s(\gamma_0 (n \Delta - \tau_0)) (n \Delta - \tau_0) \dot{s}^*(\gamma_0 (n \Delta - \tau_0)) \right )^2 \right \},
\end{align}
\begin{align}
\nonumber & \lim_{\sigma \rightarrow 0} \textstyle \mathbb{E} \left \{ b(\xi(\omega)) d(\eta(\omega)) \frac{X_1 Y_1}{\sigma^2} \right \} = -b(0) d(0) \frac{\gamma_0 \left | x \right |^2 E_s}{2} \times \\
& \textstyle \text{Re} \left \{ \left [ \sum_{n \in \mathcal{N}} s(\gamma_0 (n \Delta - \tau_0)) \dot{s}^*(\gamma_0 (n \Delta - \tau_0)) \right ] \times \right . \\
\nonumber & \textstyle \left . \left [ \sum_{n \in \mathcal{N}} s^*(\gamma_0 (n \Delta - \tau_0)) (n \Delta - \tau_0) \dot{s}(\gamma_0 (n \Delta - \tau_0)) \right ] \right \},
\end{align}
\begin{align}
\nonumber & \lim_{\sigma \rightarrow 0} \textstyle \mathbb{E} \left \{ b(\xi(\omega)) d(\eta(\omega)) \frac{X_1 Y_3}{\sigma^2} \right \} = -b(0) d(0) \frac{\gamma_0 \left | x \right |^2 E_s}{2} \times \\
& \textstyle \text{Re} \left \{ \left [ \sum_{n \in \mathcal{N}} s(\gamma_0 (n \Delta - \tau_0)) \dot{s}^*(\gamma_0 (n \Delta - \tau_0)) \right ] \times \right . \\
\nonumber & \textstyle \left . \left [ \sum_{n \in \mathcal{N}} s(\gamma_0 (n \Delta - \tau_0)) (n \Delta - \tau_0) \dot{s}^*(\gamma_0 (n \Delta - \tau_0)) \right ] \right \},
\end{align}
\begin{align}
\nonumber & \lim_{\sigma \rightarrow 0} \textstyle \mathbb{E} \left \{ b(\xi(\omega)) d(\eta(\omega)) \frac{X_3 Y_1}{\sigma^2} \right \} = -b(0) d(0) \frac{\gamma_0 \left | x \right |^2 E_s}{2} \times \\
& \textstyle \text{Re} \left \{ \left [ \sum_{n \in \mathcal{N}} s(\gamma_0 (n \Delta - \tau_0)) \dot{s}^*(\gamma_0 (n \Delta - \tau_0)) \right ] \times \right . \\
\nonumber & \textstyle \left . \left [ \sum_{n \in \mathcal{N}} s(\gamma_0 (n \Delta - \tau_0)) (n \Delta - \tau_0) \dot{s}^*(\gamma_0 (n \Delta - \tau_0)) \right ] \right \},
\end{align}
\begin{align}\label{equ:0049}
\nonumber & \lim_{\sigma \rightarrow 0} \textstyle \mathbb{E} \left \{ b(\xi(\omega)) d(\eta(\omega)) \frac{X_3 Y_3}{\sigma^2} \right \} = -b(0) d(0) \frac{\gamma_0 \left | x \right |^2 E_s^2}{2} \times \\
& \textstyle \sum_{n \in \mathcal{N}} (n \Delta - \tau_0) \left | \dot{s}(\gamma_0(n \Delta - \tau_0)) \right |^2,
\end{align}
\end{subequations}
Specifically, in deriving the results given by \eqref{equ:0048}--\eqref{equ:0049}, it is necessary to apply the relation
\begin{equation}
\text{Re} \left \{ R \right \} \text{Re} \left \{ U \right \} = \frac{1}{4} \left ( R U + R^* U + R U^* + R^* U^* \right ),
\end{equation}
which always holds for any two complex terms $R$ and $U$. Moreover, two properties of the complex Gaussian noise are also utilized \cite{Luo95, Yin98}:
\begin{subequations}
\begin{align}
& \mathbb{E} \left \{ \frac{w(n \Delta)}{\sigma} \frac{w^*(m \Delta)}{\sigma} \right \} = \left \{
\begin{array}{ll}
1, & \text{if } n = m, \\
0, & \text{if } n \neq m,
\end{array}
\right . \\
& \mathbb{E} \left \{ \frac{w(n \Delta)}{\sigma} \frac{w(m \Delta)}{\sigma} \right \} = 0, \quad \forall n, \forall m.
\end{align}
\end{subequations}

The limits of the rest terms on the RHS of \eqref{equ:045} are all shown to equal zero as $\sigma \rightarrow 0$.

\end{document}